\documentclass[amssymb,amsmath,aps,showpacs,floatfix,nofootinbib,showpacs,12pt]{revtex4}
\usepackage{bm}
\usepackage{dcolumn}
\usepackage{graphicx}
\usepackage{amsfonts}
\usepackage{amssymb}
\usepackage{latexsym}
\usepackage{amsmath}

\newcommand{\barr}{\begin{eqnarray}}
\newcommand{\earr}{\end{eqnarray}}

\begin{document}
\newcommand{\newc}{\newcommand}
\newc{\ra}{\rightarrow}
\newc{\lra}{\leftrightarrow}
\newc{\beq}{\begin{equation}}
\newc{\eeq}{\end{equation}}
\def\bea{\begin{eqnarray}}
\def\eea{\end{eqnarray}}
\def\nnb{\nonumber}
\newcommand{\half}{\frac{1}{2}}
\newcommand{\Vcc}{V_{\mathrm{CC}}}
\newcommand{\Vnc}{V_{\mathrm{NC}}}
\newcommand{\Rnc}{R_{\mathrm{NC}}}
\newcommand{\VR}{V_{\mathrm{R}}}

\newcommand{\Od}{{\cal O}}
\newcommand{\lsim}   {\mathrel{\mathop{\kern 0pt \rlap
  {\raise.2ex\hbox{$<$}}}
  \lower.9ex\hbox{\kern-.190em $\sim$}}}
\newcommand{\gsim}   {\mathrel{\mathop{\kern 0pt \rlap
  {\raise.2ex\hbox{$>$}}}
  \lower.9ex\hbox{\kern-.190em $\sim$}}}

%
%

\title {Neutrino oscillations in the presence of super-light sterile neutrinos. }

\title {Neutrino oscillations in the presence of super  light sterile neutrinos. }

\author{P.C. Divari \thanks{pdivari@gmail.com} }

\affiliation{{\it Department of Physical Sciences and
Applications, Hellenic Army Academy,\\ Vari 16673, Attica,
Greece}}

\author{J. D. Vergados \thanks{Vergados@cc.uoi.gr}}

\affiliation{Theoretical Physics Division, University of
Ioannina,\\ Ioannina, GR 451 10, Greece}

\begin{abstract}
 In the present paper we study the effect of
conversion of super-light sterile neutrino (SLSN) to electron
neutrino in matter like that of the Earth. In the Sun the
resonance conversion between SLSN and electron neutrino  via the
neutral current is suppressed due to the smallness of neutron
number. On the other hand, neutron number density can play an
important role in the Earth, making the scenario of SLSN quite
interesting. The effect of CP-violating phases on active-SLSN
oscillations is also discussed. Reactor neutrino experiments with
medium or short baseline may probe the scenario of SLSN.

\end{abstract}

\maketitle
\section{Introduction}
\label{sec1}

Neutrinos are one of the most interesting  constituents  of
particle physics. They interact only via the weak interaction and
are nearly massless. In the standard picture, there are three
neutrino species $\nu_1$, $\nu_2$ and $\nu_3$, with a summed mass
that solar and atmospheric oscillation observations bound to be
above 0.06 eV (e.g. \cite{Gonzalez,Scholberg}). Specifically, the
neutrino oscillation depends on two mass splittings (e.g. $\Delta
m_{21}^2$ and $\Delta m_{31}^2$), three mixing angles
($\theta_{12}$, $\theta_{13}$ and $\theta_{23}$) and a
CP-violating Dirac phase, $\delta_D$. In fact, the oscillation
data show that the three ordinary active neutrinos $\nu_e$,
$\nu_{\mu}$ and $\nu_{\tau}$ are mainly mixed with the three light
neutrinos $\nu_1$, $\nu_2$ and $\nu_3$ with masses such that
\begin{equation} \label{eq1a}
 \Delta m_{sol}^2=\Delta m_{21}^2
\end{equation}
\begin{equation} \label{eq1b}
 \Delta m_{atm}^2=|\Delta m_{31}^2|\simeq|\Delta m_{32}^2|
\end{equation}
with $\Delta m_{ij}^2\equiv m_i^2-m_j^2$. $\Delta m_{atm}^2$ and
$\Delta m_{sol}^2$ stands for the atmospheric and the solar
mass-squared splitting, respectively.

\begin{figure}[htb]
\centering
\includegraphics[width=.4\textheight]{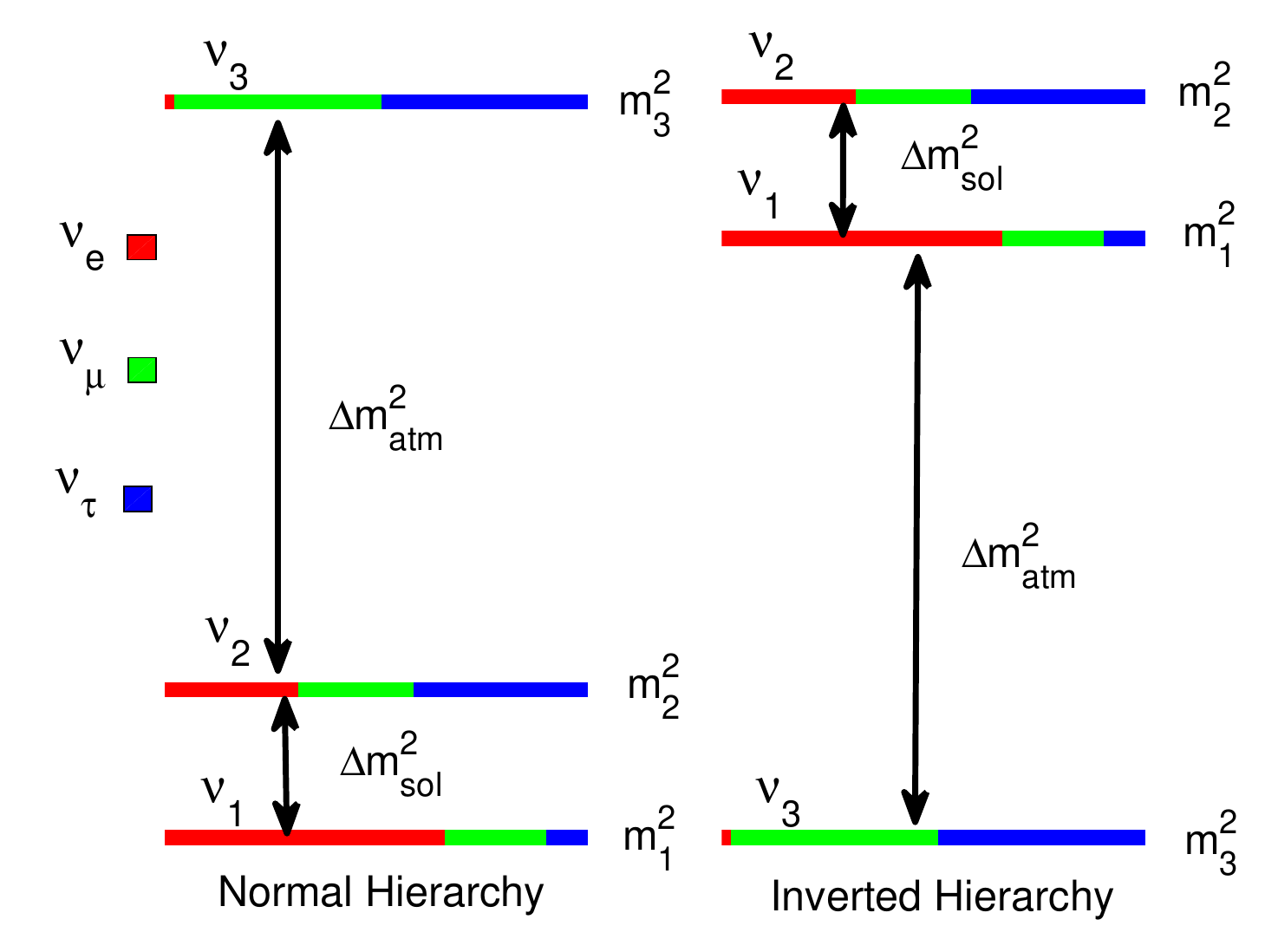}
\caption{(Color on line). Pattern of neutrino masses for the
normal and inverted hierarchies is shown as mass squared. Flavor
composition is indicated by the horizontal divisions  ($\nu_{e}$
on the left, $\nu_{\mu}$ in the middle, and $\nu_{\tau}$ on the
right).
 $\Delta m_{atm}^2=|\Delta
m_{31}^2|\simeq|\Delta m_{32}^2|$ and $\Delta m_{sol}^2=\Delta
m_{21}^2$ stands for the atmospheric and the solar mass-squared
splitting, respectively.}
 \label{hier}
  \end{figure}

The neutrino mass hierarchy (or the ordering of the neutrino
masses), i.e., whether the $\nu_3$ neutrino mass eigenstate is
heavier or lighter than the $\nu_1$ and $\nu_2$ mass eigenstates,
is one of the remaining undetermined fundamental features of the
neutrino Standard Model.
 The scenario, in which the $\nu_3$ is heavier,
is referred to as the normal mass hierarchy (NH). The other
scenario, in which the $\nu_3$  is lighter, is referred to as the
inverted mass hierarchy (IH). The pattern of neutrino masses and
mixings is schematically shown in Fig.\hspace{2pt}\ref{hier}.
 Most of  the neutrino parameters entering neutrino oscillation
formula are well determined  except for the CP violating phase
$\delta_D$ and the sign of $m_{3}^2-m_{1}^2$ (the neutrino mass
hierarchy pattern). From all the recent sensitivity studies it has
clearly emerged that, at least for the next five years, it will be
extremely difficult  for a single experiment to provide definitive
information for any of the two searched properties CP-violation
and neutrino-mass-hierarchy.

The neutrino sector may be richer than commonly believed and not
confined to the 3-flavor framework. Several anomalies have
recently emerged in short base line (SBL) oscillation experiments,
which indicate significant extensions  in the Standard picture and
point towards the existence of new physics. The most famous
examples are the $\nu_{\mu}\rightarrow \nu_{e}$ and/or
$\bar\nu_{\mu}\rightarrow \bar\nu_{e}$ transitions in short
baseline LSND and MiniBooNE experiments
\cite{MiniBooNE,MiniBooNE2,MiniBooNE3}, reactor neutrino deficit
\cite{Mueller} and Gallium anomaly \cite{Abdurashitov,Kopp}.
A recent careful analysis of neutrino anomaly (NA)
\cite{Giunti,Mention,KoppPRL} led to a challenging  suggestion
that there may be one or more additional eV scale massive sterile
neutrinos \cite{v1,v2,v3,v4,v5,v6,v7,Pilaftsis}. One extra sterile
neutrino is also suggested by recent analysis of the data from
cosmological observations and Big-Bang Nucleosynthesis, in order
to explain the existence of additional dark radiation in the
Universe
\cite{Hamann,Hamann2,Hamann3,Hamann4,Archidiacono,Mangano,Mangano2,Mangano3,Riess,Kusenko,Wyman,Battye}.

Furthermore, the large mixing angle (LMA)
Mikheyev-Smirnov-Wolfenstein (MSW) solution
\cite{Wolfenstein,Mikheyev,Mikheyev2,Mikheyev3} to the solar
neutrino problem predicts  an upturn of the energy spectrum of
events at energies below 8 MeV. However, recent measurements of
the energy spectra of the solar neutrino events at
Super-Kamiokande-III \cite{Super}, SNO \cite{SNO}, and Borexino
\cite{Borexino} experiments do not show the expected (according to
LMA) upturns at low energies. In \cite{Holanda04,Holanda11}, a
scenario is proposed to explain the suppression of the upturn. The
scenario is based on the possible existence of  a  super-light
sterile neutrino (SLSN) which weakly mixes with the active
neutrinos. The new mass eigenstate is called $\nu_0$ and its mass
is denoted by $m_0$. To explain the suppression of the upturn in
the low energy solar data, it is shown that the mass squared
difference with mass state $\nu_1$ is around $\Delta
m_{01}^2\approx(0.7-2)\times 10^{-5}eV^2$ and the mixing angle
with electron neutrino around $sin^22\theta_{01}\approx
(0.001-0.005)$ \cite{Holanda04,Holanda11}. Such a mixing leads to
appearance of a dip in the $\nu_e$- survival probability in the
energy range (1 - 7) MeV, thus removing the upturn of the spectra.
This is achieved with the help of a MSW resonant conversion of
this  SLSN with solar electron neutrino when neutrino travels from
the interior of the Sun to the outside.

One of the major concerns in neutrino oscillation experiments is
the effect of Earth matter in neutrino flavor conversion. One
would naively expect that a similar  resonant flavor conversion
between SLSN  and electron neutrino should also happen when
neutrinos propagate in Earth matter. In the Sun   the neutron
number density is small, thus the effect of neutral current
interaction $V_n$ can be effectively neglected. On the other hand,
in Earth matter the neutron number density is of the same order of
magnitude with
 electron number density \cite{Kim,PREM,Lisi}. Therefore,
the effect of $V_n$  can play an important role.

 In the present paper,  we present  a comprehensive study of  the
dependence of SLSN transition probability   on a sizeable
potential of the neutral current interaction with matter. We
consider a four-neutrino ($4\nu$)  scheme to calculate active-SLSN
oscillations in matter, extending the study presented in Ref.
\cite{WLiao} where it was assumed that the electron neutrino has
non-negligible mixing only with the two massive neutrinos which
generate the solar squared-mass difference. The effect of
CP-violating phases is also included.

 This paper is organized as follows. In Section \ref{sec2}
the $4\nu$ framework of neutrino evolution in Earth matter is
given, adopting a simplified version of Earth reference model. In
Section \ref{sec3} we give numerical results for the conversion
probability of electron neutrino to sterile one, in terms of
mixing and mass splitting parameters. Effects of the CP violation
to the survival and transition probabilities   are illustrated in
Section \ref{sec3a}. For the completeness  of the discussion in
this article we explore the capacity of a reactor experiment to
probe the super-light sterile neutrino scenario (Sec. \ref{sec4}).
Conclusions are drawn in Section. \ref{sec5}.

\section{Four neutrino scheme }
\label{sec2}

The four neutrino mixing matrix can be described by six mixing
angles and three physical Dirac phases. If the neutrinos are of
Majorana type, there will  also be three Majorana CP-violating
phases which do not show up in the neutrino oscillation patterns.
Following the notation in \cite{Holanda11}, we call the mass
eigenstates as $(\nu_0,\nu_1,\nu_2,\nu_3)$ with mass eigenvalues
$(m_0,m_1,m_2,m_3)$. The flavor eigenstates are related to mass
eigenstates by a $4\times 4$ unitary matrix, $U$ as follows
\begin{eqnarray}
 \begin{pmatrix} \nu_s \cr \nu_e \cr \nu_\mu \cr \nu_\tau
\end{pmatrix} =U\cdot \begin{pmatrix} \nu_0 \cr \nu_1 \cr \nu_2 \cr
\nu_3
\end{pmatrix}
\end{eqnarray}
The sterile neutrino, $\nu_s$, is mainly present in the mass
eigenstate $\nu_0$  with mass $m_0$. It mixes weakly with active
neutrinos and this mixing can be treated as a small perturbation
of the standard LMA structure.
The matrix $U$ is a $4\times 4$ unitary matrix describing the
mixing of neutrinos. Neglecting CP violating phases, it can be
parameterized by
\begin{eqnarray}
U=R(\theta_{23}) R(\theta_{13}) R(\theta_{12}) R(\theta_{02})
R(\theta_{01}) R(\theta_{03}) \label{H0c},
\end{eqnarray}
where $R(\theta_{ij})$ is a $4\times 4$ rotation matrix with a
mixing angle $\theta_{ij}$ appearing at $i$ and $j$ entries, e.g.
\bea R(\theta_{01})=\begin{pmatrix} \cos\theta_{01} &
\sin\theta_{01} & 0 & 0 \cr -\sin\theta_{01} & \cos\theta_{01} & 0
& 0 \cr
 0 & 0 & 1 & 0 \cr 0 & 0 & 0 & 1 \cr \end{pmatrix}, \quad R(\theta_{13})=\begin{pmatrix} 1 &
0 & 0 & 0 \cr 0 & \cos\theta_{13} & 0 & \sin\theta_{13} \cr
 0 & 0 & 1 & 0 \cr 0 & -\sin\theta_{13} & 0 & \cos\theta_{13} \cr \end{pmatrix},\label{R01}
\end{eqnarray}
\begin{eqnarray} R(\theta_{02})=\begin{pmatrix} \cos\theta_{02} & 0&
\sin\theta_{02} & 0 \cr
 0 & 1 & 0 & 0 \cr
 -\sin\theta_{02} & 0 & \cos\theta_{02} & 0  \cr
 0 & 0 & 0 & 1 \cr \end{pmatrix}, \quad R(\theta_{12})=\begin{pmatrix} 1 & 0&
0 & 0 \cr
 0 & \cos\theta_{12} & \sin\theta_{12} & 0 \cr
 0& -\sin\theta_{12} & \cos\theta_{12} & 0  \cr
 0 & 0 & 0 & 1 \cr \end{pmatrix}, \label{R02}
\end{eqnarray}
 and
 \begin{eqnarray} R(\theta_{03})=\begin{pmatrix} \cos\theta_{03} & 0&
0 & \sin\theta_{03} \cr
 0 & 1 & 0 & 0 \cr
 0 & 0 & 1 & 0  \cr
 -\sin\theta_{03} & 0 & 0 & \cos\theta_{03} \cr \end{pmatrix}, \quad  R(\theta_{23})=\begin{pmatrix} 1 & 0&
0 & 0 \cr
 0 & \cos\theta_{23} & \sin\theta_{23} & 0 \cr
 0 & -\sin\theta_{23} & \cos\theta_{23} & 0  \cr
 0 & 0 & 0 & 1 \cr \end{pmatrix}.\label{R03}
\end{eqnarray}
$\theta_{12,13,23}$ are the mixing angles governing the flavor
conversion of solar neutrinos, reactor neutrinos at short baseline
and atmospheric neutrinos separately and they have been measured
in solar, atmospheric, long baseline and reactor neutrino
experiments \cite{RPP,theta13,theta13-2,theta13-3}.

\begin{figure}[htb]
\centering
\includegraphics[width=.18\textheight]{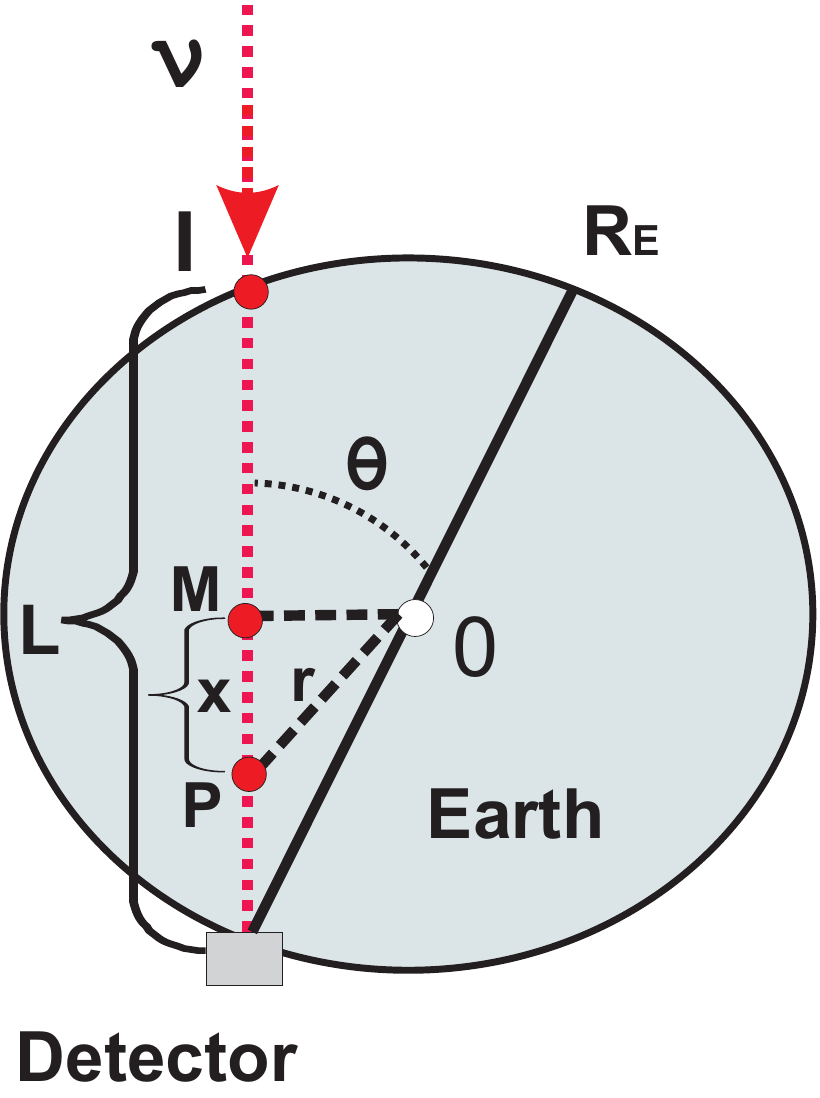}a)
\includegraphics[width=.3\textheight]{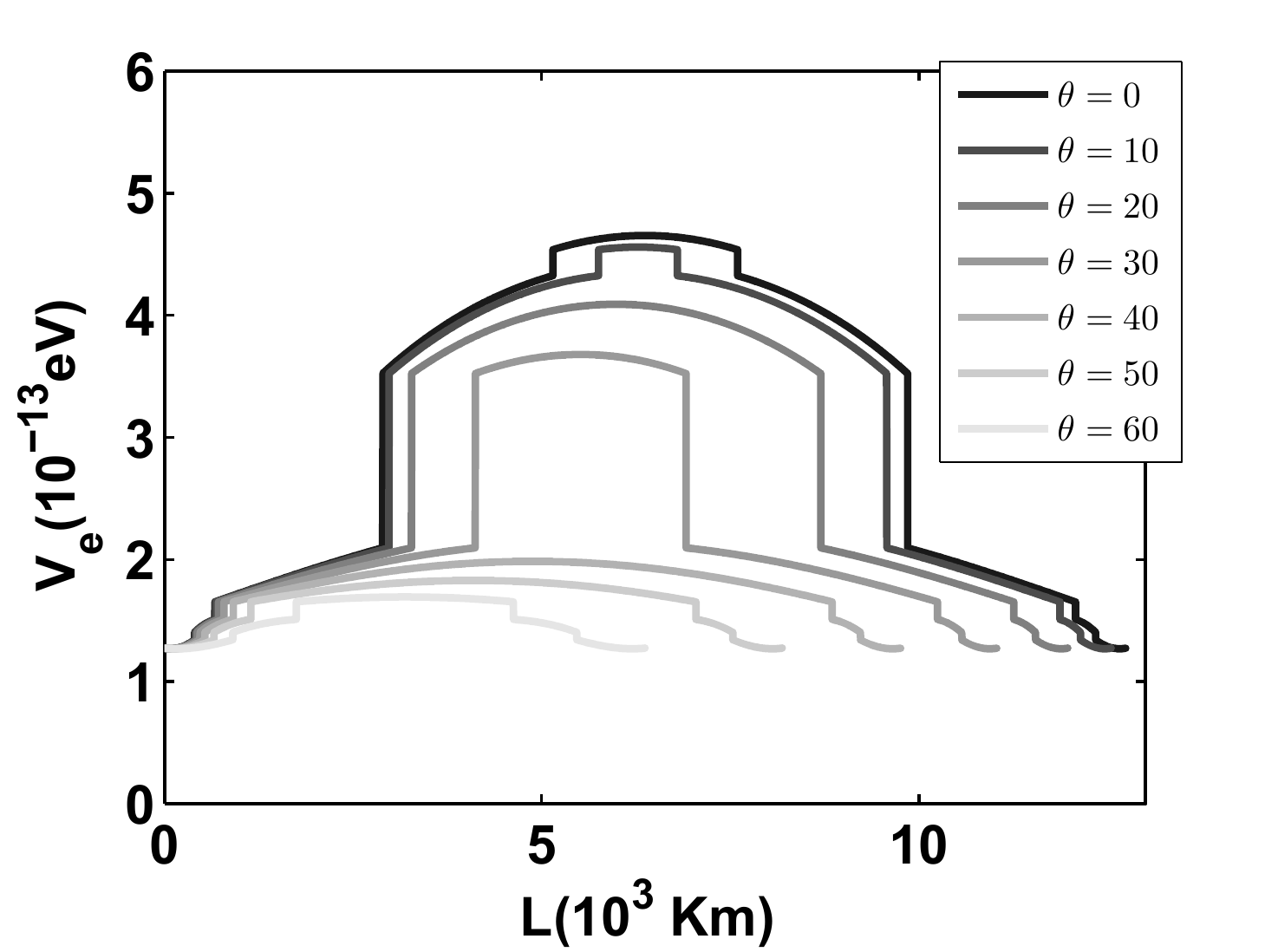}b)
\caption{(Color on line). Left panel: a schematic view of the
 solar neutrinos trajectories from the entry point (I) to the   detector.
  $\theta$ is the nadir angle of the neutrinos.  Right
panel: The electron current matter potential $V_e$  is shown as a
function of neutrino travelling distance L for   nadir angles
$\theta=0^0$ to  $\theta=60^0$. The data is based on the
Preliminary Reference Earth Model (PREM) \protect\cite{PREM}. The
Earth radius is taken $R_E=6370 Km$.}
 \label{density}
  \end{figure}

The evolution of solar neutrinos propagating in Earth matter is
described by the equation
\begin{equation}
i\frac{d\Psi}{dt}=(U{H_0}U^\dagger +{\cal V})\Psi \label{eq1}
\end{equation}
where $\Psi=(\psi_s,\psi_e,\psi_{\mu},\psi_{\tau})^T$ is the
flavour transition amplitudes and
\begin{eqnarray}
 && H_0= \frac{1}{2E}
\mbox{diag}\{\Delta m^2_{01},0,\Delta m^2_{21},\Delta m^2_{31} \}
\label{H0a} \nonumber \\
&& V=\mbox{diag} \{0, V_e+V_n,V_n,V_n \} \label{H0b}
\end{eqnarray}
where $E$ is the neutrino energy and $ \Delta{m}^2_{kj} = m_{k}^2
- m_{j}^2\,$. The charged-current and neutral-current matter
potentials are defined as
\begin{equation}
  V_{e}=\sqrt{2}G_{\mathrm{F}}N_{e} \simeq 7.63 \times 10^{-14}
  \, \frac{ N_{e} }{ N_{\mathrm{A}} \,  } \, \mathrm{eV}
  \quad, \qquad V_{n}=-\half\sqrt{2}G_{\mathrm{F}}N_{n}
  \quad,
\end{equation}
where $G_{\mathrm{F}}$ is the Fermi constant, $N_{e}$ is the
electron number density, $N_{n}$ is the neutron number density,
and $N_{\mathrm{A}}$ is the Avogadro's number.
All  neutrino flavors interact with Earth matter constituents
(electrons and neutrons) as they travel to the detection point
(see Fig.\hspace{2pt}\ref{density}). The charge-current neutrino
matter potential   used in this paper is taken from Ref.
\cite{PREM}. Since in the Earth the neutron number density is
roughly of the same order of the electron number density,  the
neutral-current is taken to be $V_n=-0.5V_e$.

\section{Numerical Results}
\label{sec3}

In order to calculate the neutrino evolution inside the Earth
matter,  an accurate description of the Earth density profile   is
needed. For this reason    a simplified version of the preliminary
Earth reference model (PREM) \cite{PREM} is employed , which
contains  five shells \cite{Lisi} and uses the polynomial function
\begin{equation}
  N_{ei}(r)=(\alpha_i + \beta_i\,r^2 + \gamma_i\,r^4)N_A
  \label{rd}
\end{equation}
for the $i$-th shell ($1\leqslant i\leqslant 5$, where $i=1$ is
the innermost shell) to describe the Earth's electron density at
the radial distance $r$  (see Fig.\hspace{2pt}\ref{density}(a)).
The values of the coefficients are given in
Table~\ref{tb:Ne_earth} for nadir angle $\theta=0$. For
  nadir angles $\theta\neq 0$ Eq. (\ref{rd})  becomes
  \begin{equation}
  N_{ei}(r)=(\alpha_i' + \beta_i'\,x^2 + \gamma_i'\,x^4)N_A
  \label{rdn}
\end{equation}
where
\begin{eqnarray}
 && \alpha_i'=\alpha_i + \beta_i\,sin^2\theta + \gamma_i\,sin^4\theta
\label{cofa} \nonumber \\
&&  \beta_i'= \beta_i  + 2\gamma_i\,sin^2\theta \label{cofb} \nonumber \\
&&  \gamma_i'=  \gamma_i \label{cofc}
\end{eqnarray}
where $x$ is the distance from the trajectory midpoint M to the
generic position of the neutrino (see
Fig.\hspace{2pt}\ref{density}(a)).
\begin{table}[ph]
\caption{Descriptions of the simplified PREM model with five
shells. The shell names and the values of the coefficients are
quoted from Table 1 of Ref.~\protect\cite{Lisi} (see text for
details). The radial distance $r$ is normalized to the Earth
radius $R_E$.} {\begin{tabular}{@{}cccccc@{}} \hline\hline $i$ &
{\text{Shell}} & $[r_{i-1},\,r_i]$  &
$\alpha_i$ & $\beta_i$ & $\gamma_i$ \\
\hline\hline
1 & Inner core      &     $[0,\,0.192]$        &   6.099 & $-$4.119 &    0.000   \\
2 & Outer core      &   $[0.192,\,0.546]$      &   5.803 & $-$3.653 & $-$1.086   \\
3 & Lower mantle    &   $[0.546,\,0.895]$      &   3.156 & $-$1.459 &    0.280   \\
4 & Transition Zone &   $[0.895,\,0.937]$      &$-$5.376 &   19.210 &$-$12.520   \\
5 & Upper mantle    &     $[0.937,\,1]$        &  11.540 &$-$20.280 &   10.410  \\
\hline\hline
\end{tabular} \label{tb:Ne_earth}}
\end{table}

Since, it is rather difficult to study equation Eq.~(\ref{eq1})
analytically,  a numerical treatment based on the fourth-order
Runge-Kutta method is used to solve the evolution equation of the
neutrino states. Relevant neutrino parameters in this calculation
are~\cite{RPP} \bea & \Delta m^2_{21}=(7.50\pm 0.20) \times
10^{-5} ~\textrm{eV}^2,~~
|\Delta m^2_{31}|=(2.32^{+0.12}_{-0.08})\times 10^{-3} ~\textrm{eV}^2, \label{neuparam0} \\
& \sin^2 2\theta_{12}=0.857\pm 0.024, ~~\sin^2 2\theta_{23}>0.95.
\label{neuparam1}
 \eea
 After the discovery of
$\theta_{13}$ by Daya-Bay collaboration \cite{theta13}, confirmed
by RENO experiment~\cite{theta13-2}, a precise measurement of
$\theta_{13}$ has been achieved by Daya-Bay
experiment~\cite{theta13-3}: \bea \sin^22\theta_{13}=0.089\pm
0.010\pm 0.005 . \label{neuparam2} \eea
\begin{figure}[htb]
\centering
\includegraphics[width=.3\textheight]{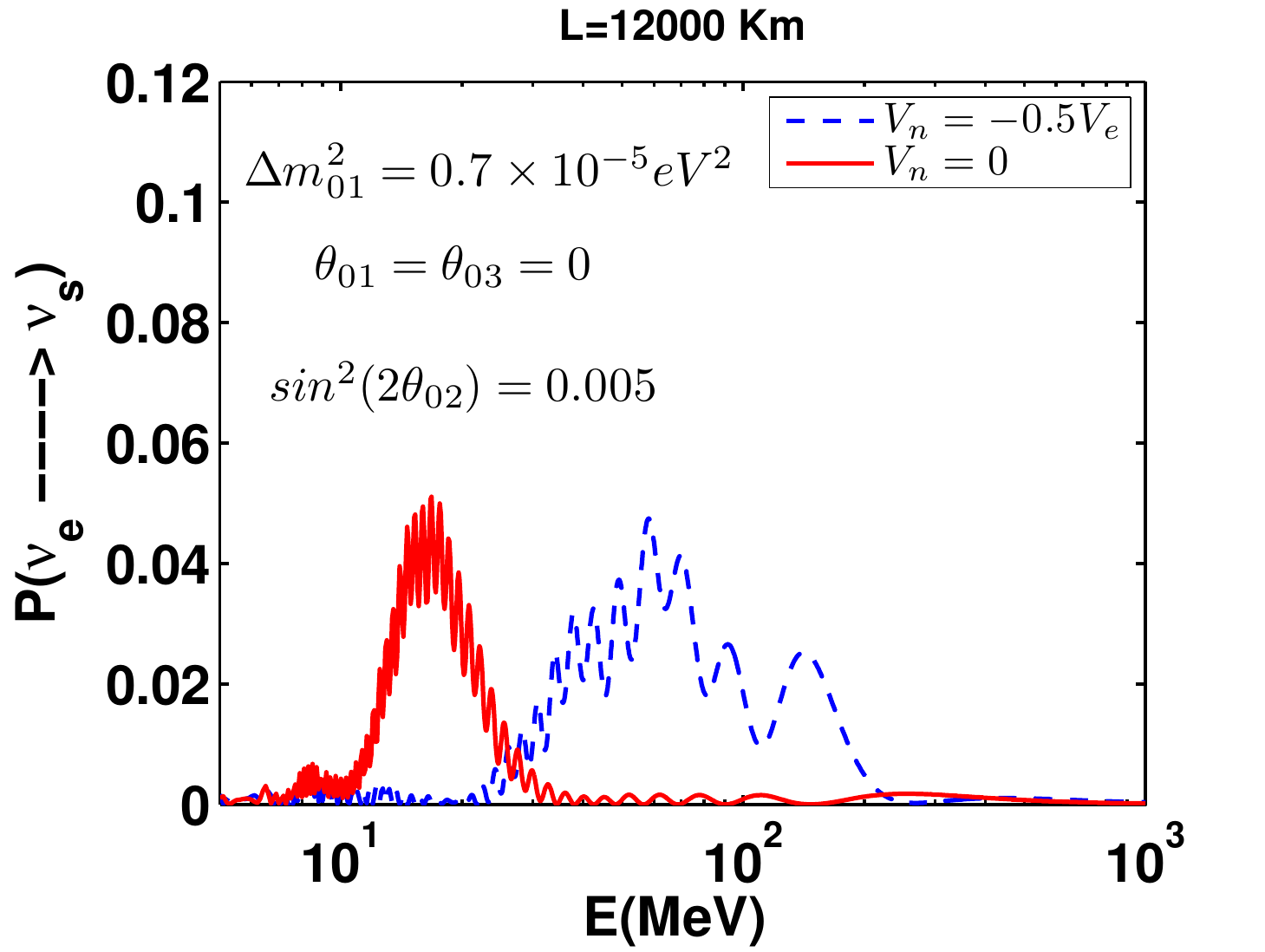}a)
\includegraphics[width=.3\textheight]{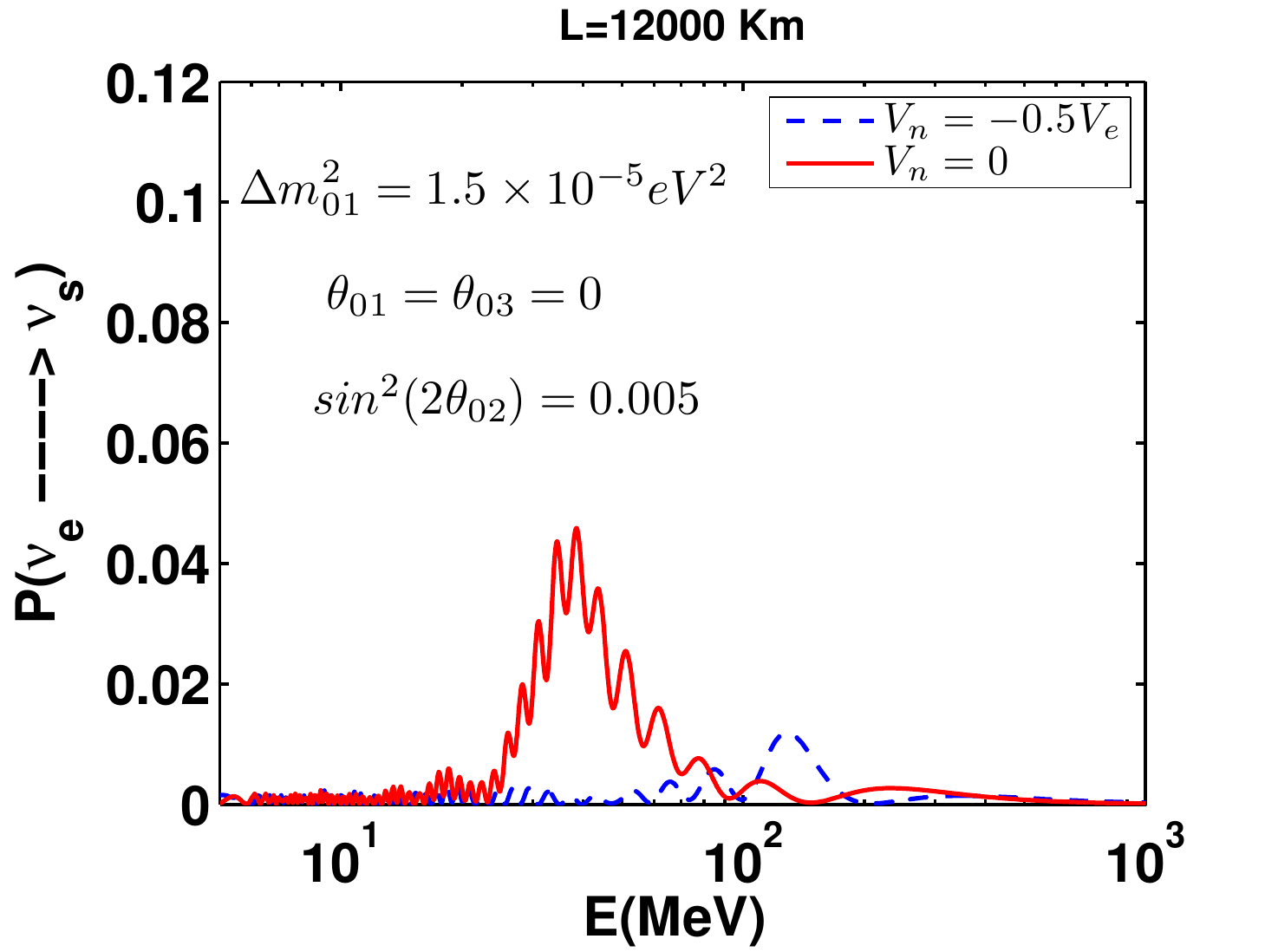}b)
\includegraphics[width=.3\textheight]{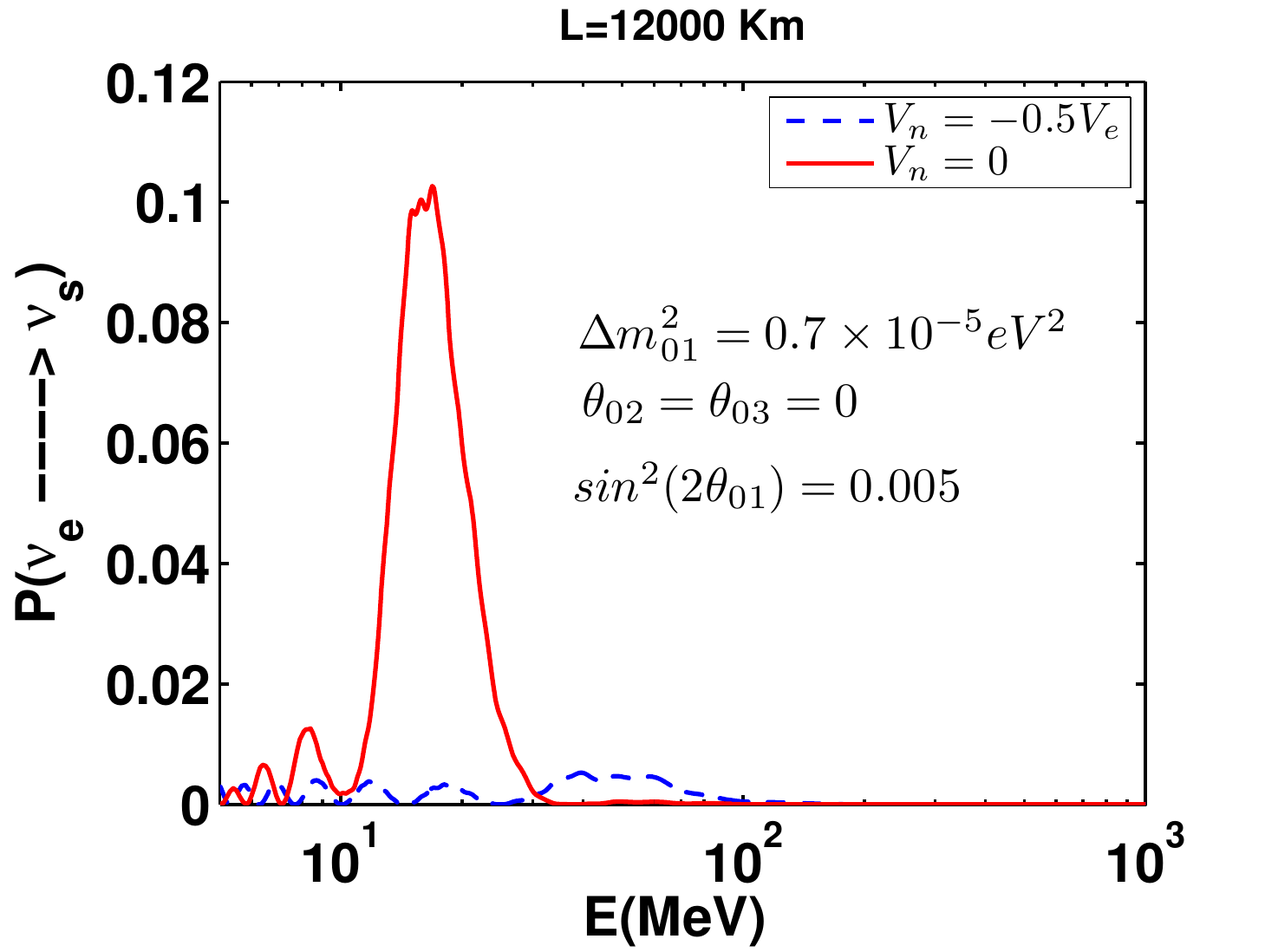}c)
\includegraphics[width=.3\textheight]{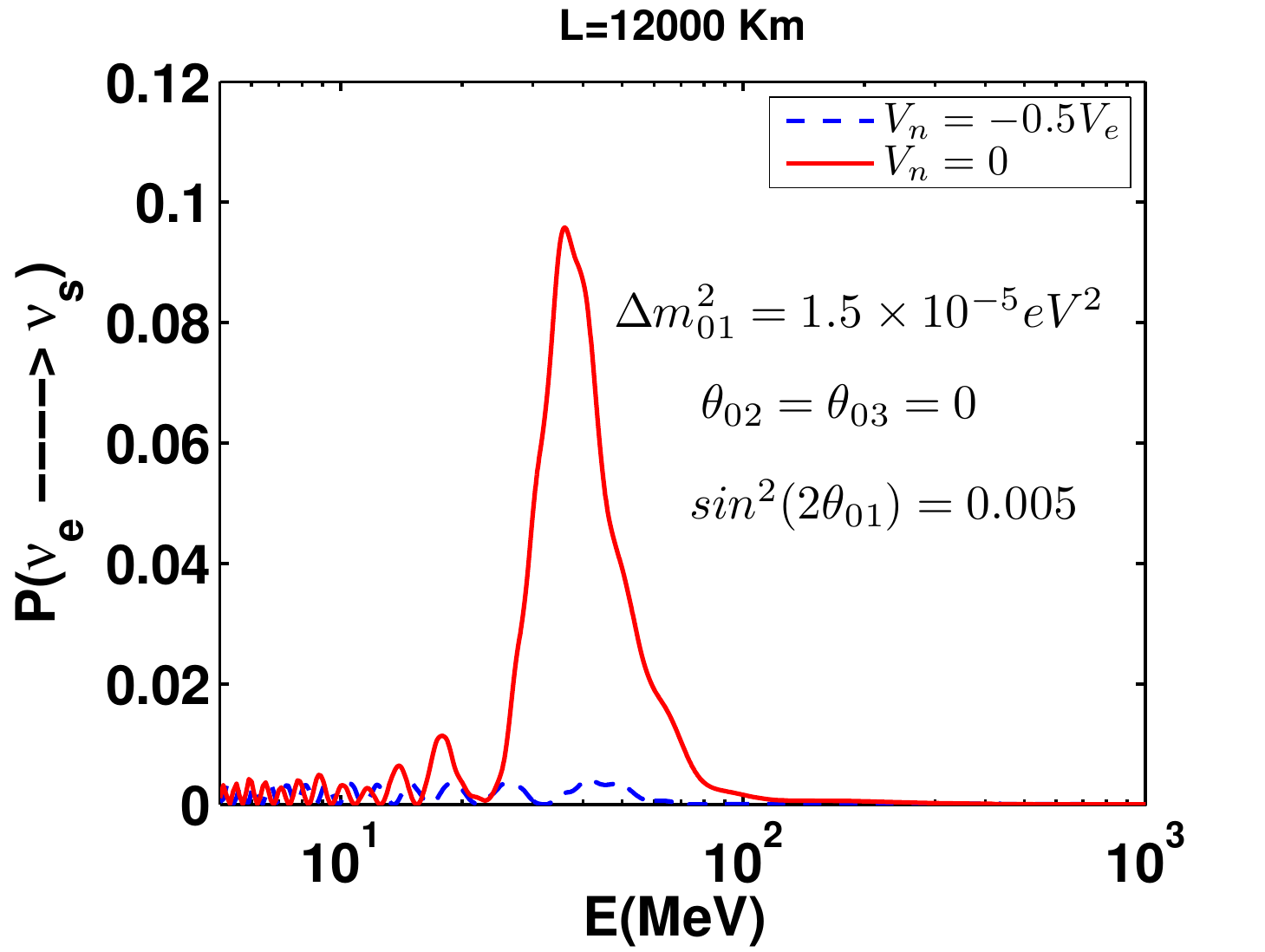}d)
\includegraphics[width=.3\textheight]{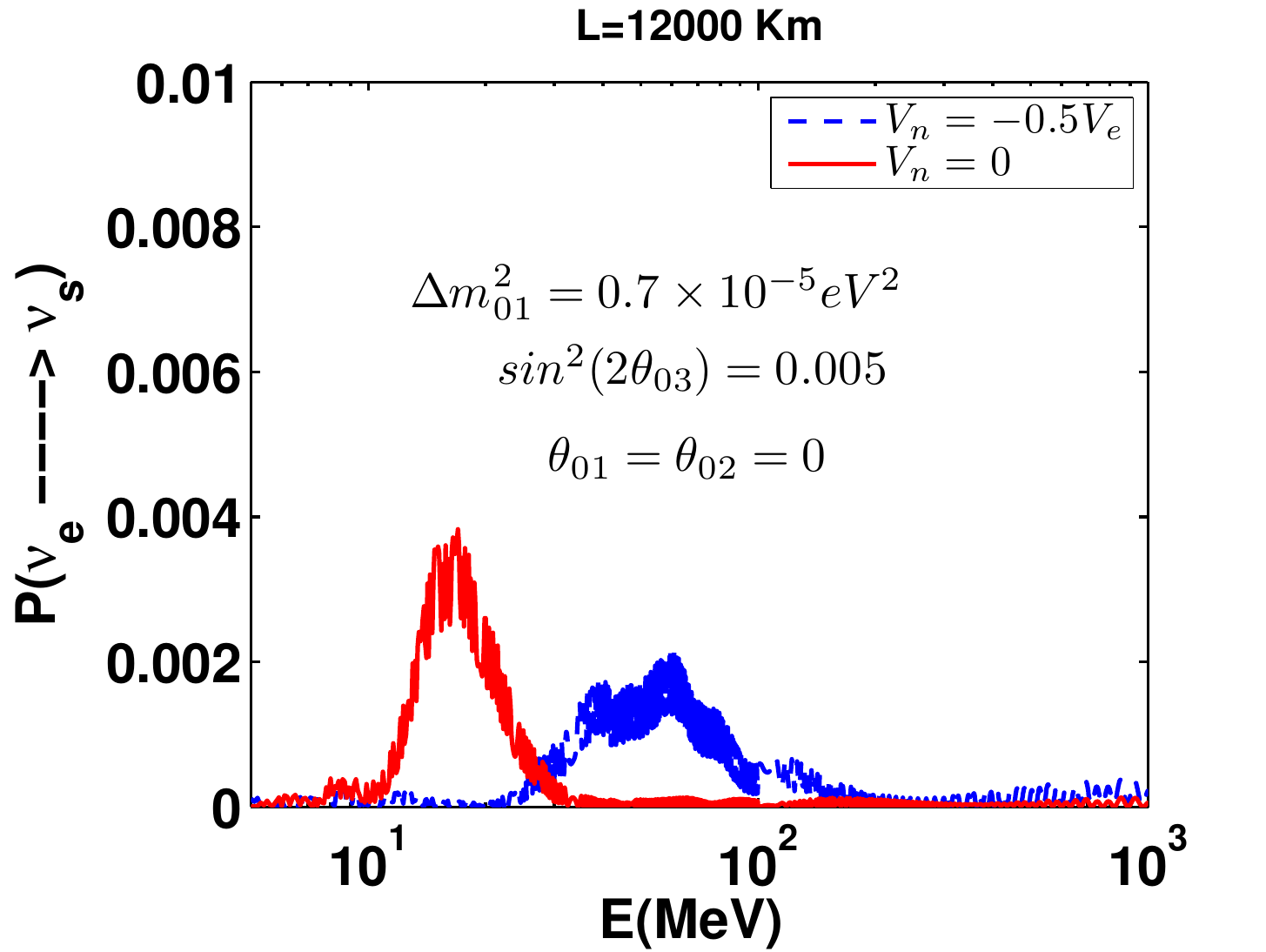}e)
\includegraphics[width=.3\textheight]{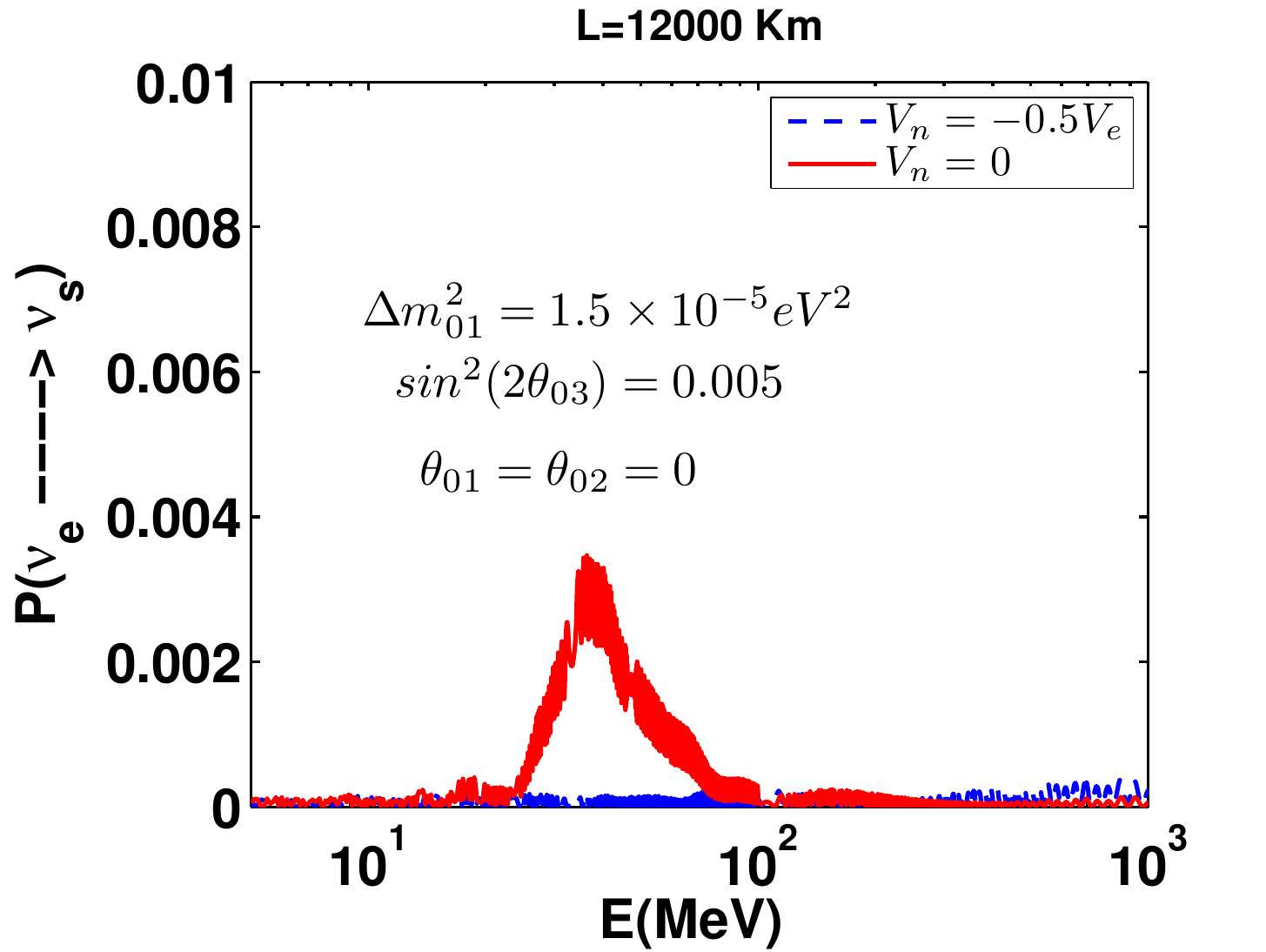}f)
\caption{(Color on line)  Conversion probability $\nu_e
\rightarrow \nu_s$ versus energy   for $V_n=0$   and for
$V_n=-0.5V_e$ at fixed   $L=12000$ km. }
 \label{cv}
  \end{figure}

The future reactor neutrino experiments of JUNO and RENO-50
\cite{Bakhti}, which are mainly proposed to determine the neutrino
mass hierarchy, predict that after 20 years of data the bounds on
$\sin^2 \theta_{01}$ and $\sin^2 \theta_{02}$ at $\Delta
m_{01}^2=2\times 10^{-5}$ eV$^2$ will be at best down to
$2.8\times 10^{-3}$ and $4.2\times 10^{-3}$, respectively. These
values are laying inside the parameter range indicated in
\cite{Holanda04,Holanda11} ({\it i.e.,} $\Delta
m_{01}^2=(0.7-2)\times 10^{-5}$~eV$^2$ and
$\sin^2\theta_{01},\sin^2 \theta_{02}\sim 10^{-3}$). For
intermediate baselines for which $ \Delta
 m_{31}^2 L/E\sim \pi$, $\theta_{01}$ and $\theta_{02}$ parameters cannot be
 resolved. 
 The atmospheric data  \cite{Cirelli} and MINOS  experiment \cite{MINOS}
have  already put the constrain
 \begin{equation}
 \label{MIN-at}
 \sin^2 \theta_{03}<0.2
 \end{equation}
 More stringent bounds are placing   from  cosmology. Recent PLANCK
 data   constrain   the effective number of relativistic  species $N_{eff}$, before the big bang
 nucleosynthesis  (BBN) epoch \cite{Ade}. If    $\theta_{03}$ or $\theta_{02}$ are large enough, $\nu_s$
  can  reach thermal equilibrium at the early universe and   can be
considered as an extra degree of freedom, contributing to
 $N_{eff}=3.68^{+0.80}_{-0.70}$ \cite{Izotov}.
 From this
 observation, Ref. \cite{Mirizzi} puts  stronger bounds
 $$\sin^2 \theta_{01}, \sin^2 \theta_{02},\sin^2 \theta_{03}<10^{-3}$$
In the following analysis   the  dependence  of mixing angle
$\theta_{01}$, $\theta_{02}$ and $\theta_{03}$ parameters as well
as the splitting mass $\Delta { m}^2_{01}$  parameter on neutrino
oscillations are investigated.

Fig.\hspace{2pt}\ref{cv} depicts the conversion probability
$\nu_{e} \rightarrow \nu_s$ in matter versus neutrino energy $E$,
at fixed source-detector  distance $L=12000$Km, for given neutrino
masses and mixing angles \cite{RPP,theta13,theta13-2,theta13-3}.
By inspection of Fig.\hspace{2pt}\ref{cv}  the  following
observations should be made: i) The resonant conversion
probability $\nu_{e} \rightarrow \nu_s$  is much stronger when
$V_n$ is switched off ( $V_n=0$), and this happens at the energy
  around 17 MeV for $\Delta { m}^2_{01}=0.7\times 10^{-5}$
eV$^2$, and around 37 MeV when $\Delta { m}^2_{01}$ becomes
$1.5\times 10^{-5}$ eV$^2$. ii) The resonant conversion
probability decreases for $V_n$ included ($V_n=-0.5V_e$). iii)
When  $V_n$ is taken into account, the maximal conversion
probability, around 5\%, happens for $\sin^22\theta_{02}=0.005$,
$\theta_{01}=\theta_{03}=0$ with mass splitting $\Delta {
m}^2_{01}=0.7\times 10^{-5}$ eV$^2$ at      energy around 60 MeV,
which is well beyond the solar and supernovae neutrino spectrum.
iv) In case where $\sin^22\theta_{03}$ has been involved, the
conversion probability amplitude   becomes  much smaller,
maximally around 0.2\%. Ultimately, matter oscillations disappear
for $\Delta { m}^2_{01}=1.5\times 10^{-5}$ eV$^2$
(Fig.\hspace{2pt}\ref{cv}f). iv) As $\Delta {m}^2_{01} $ increases
from $0.7\times 10^{-5}$ eV$^2$ to $1.5\times 10^{-5}$ eV$^2$, the
conversion probability amplitude is strongly suppressed.

\begin{figure}[htb]
\centering
\includegraphics[width=.3\textheight]{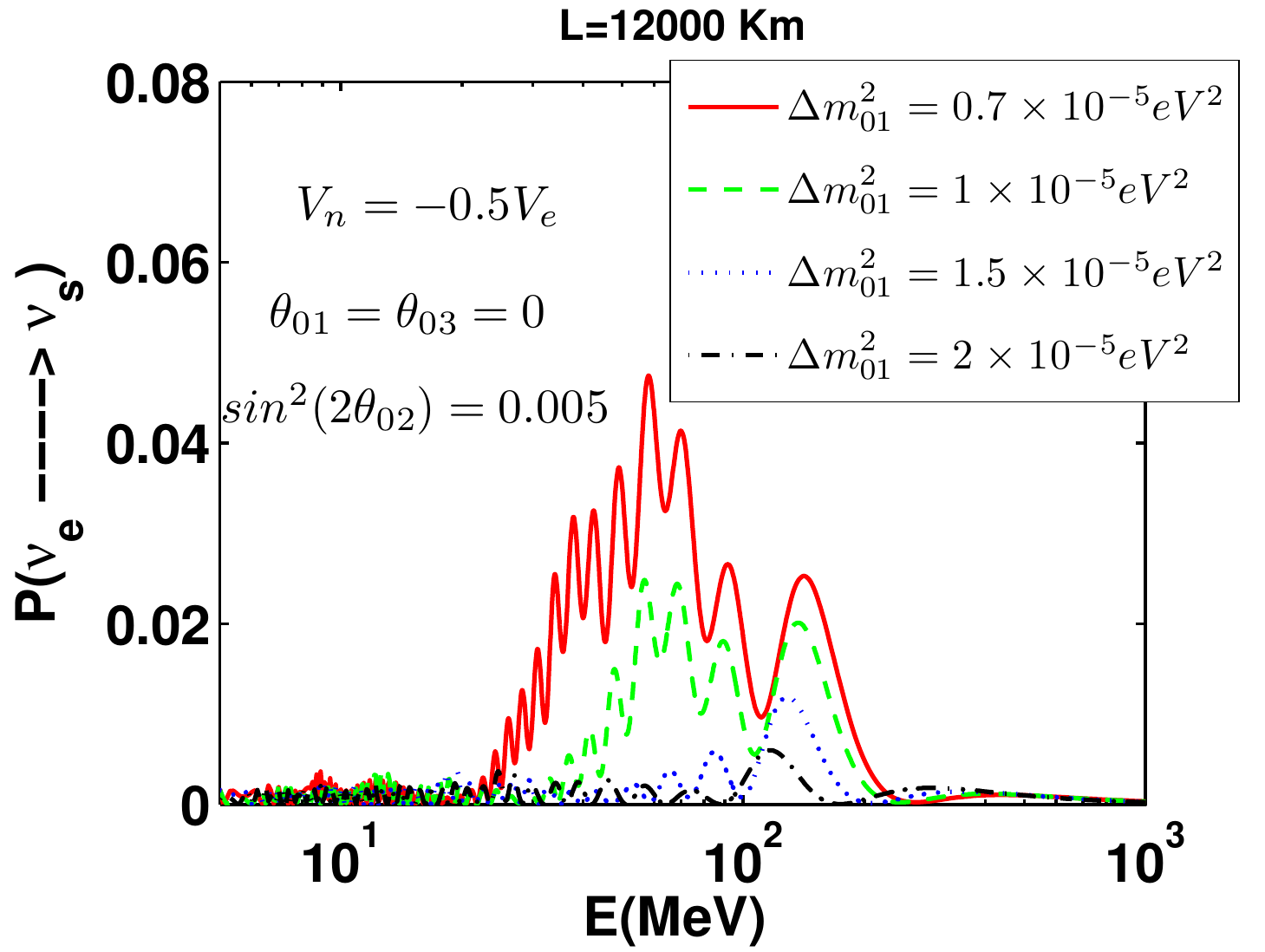}a)
\includegraphics[width=.3\textheight]{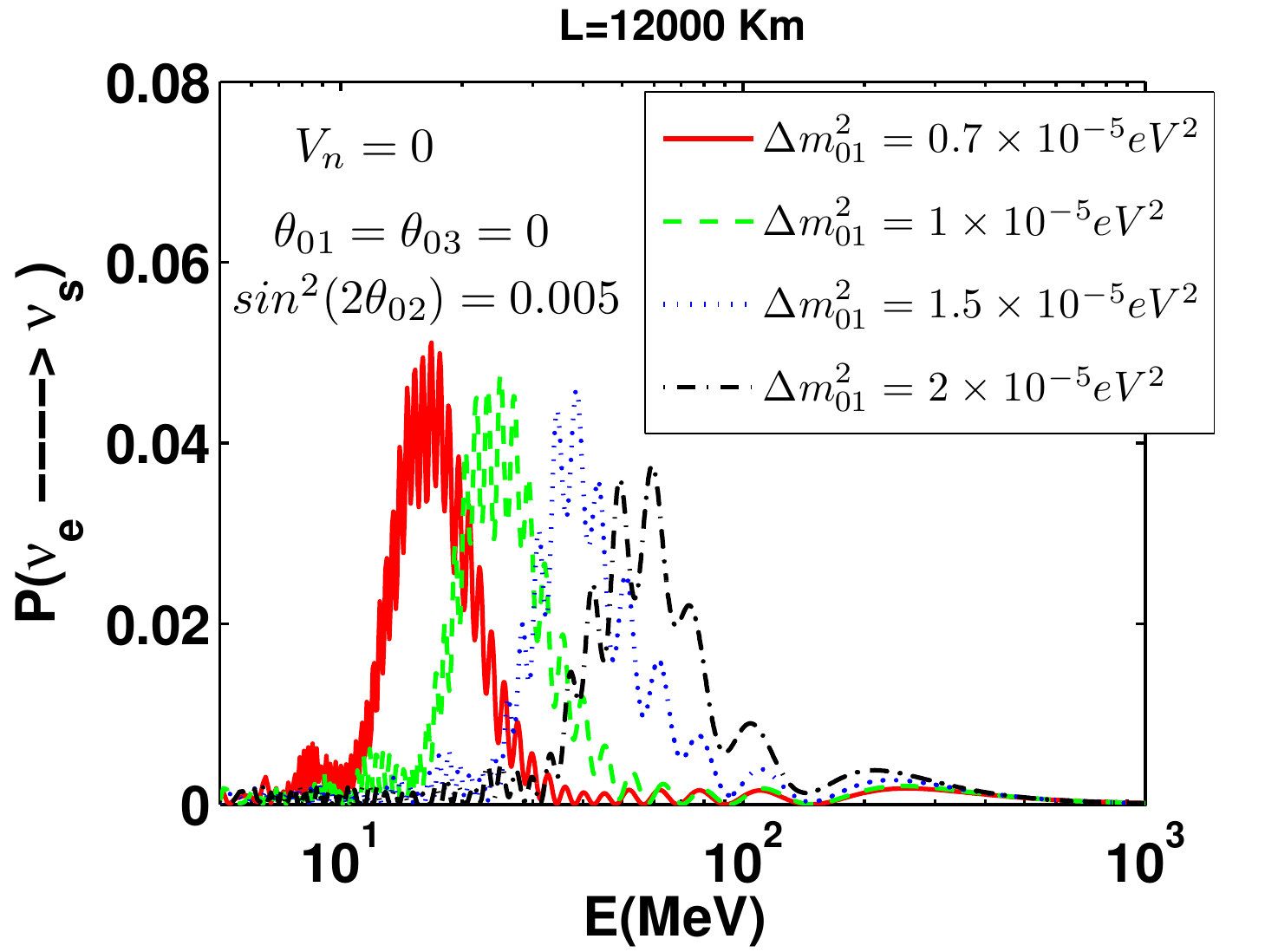}b)
\caption{(Color on line) (a) $\nu_e\rightarrow\nu_s$ conversion
probabilities versus energy  which correspond to cases with
$\Delta { m}^2_{01}=0.7 \times 10^{-5}$ eV$^{2}$, $\Delta {
m}^2_{01}=1.0 \times 10^{-5}$ eV$^{2}$,  $\Delta { m}^2_{01}=1.5
\times 10^{-5}$ eV$^{2}$ and $\Delta { m}^2_{01}=2.0 \times
10^{-5}$ eV$^{2}$ separately. (a) $V_n=-0.5V_e$ (b) $V_n=0$.
$L=12000$ Km, $\theta_{01}=\theta_{03}=0$ and $\sin^2
2\theta_{02}=0.005$ for both cases.}
 \label{mass}
  \end{figure}

More details about the variation of the conversion probability
$\nu_{e} \rightarrow \nu_s$ with respect to $\Delta {m}^2_{01}$
are given in Fig.\hspace{2pt}\ref{mass}. It is seen that as
$\Delta {m}^2_{01}$ increases the amplitude of the conversion
probability decreases. The decrease is  more rapid when $V_n$
included. Furthermore, the resonance position shifts  to greater
energies with greater mass splitting  $\Delta {m}^2_{01}$.

\begin{figure}[htb]
\centering
\includegraphics[width=.3\textheight]{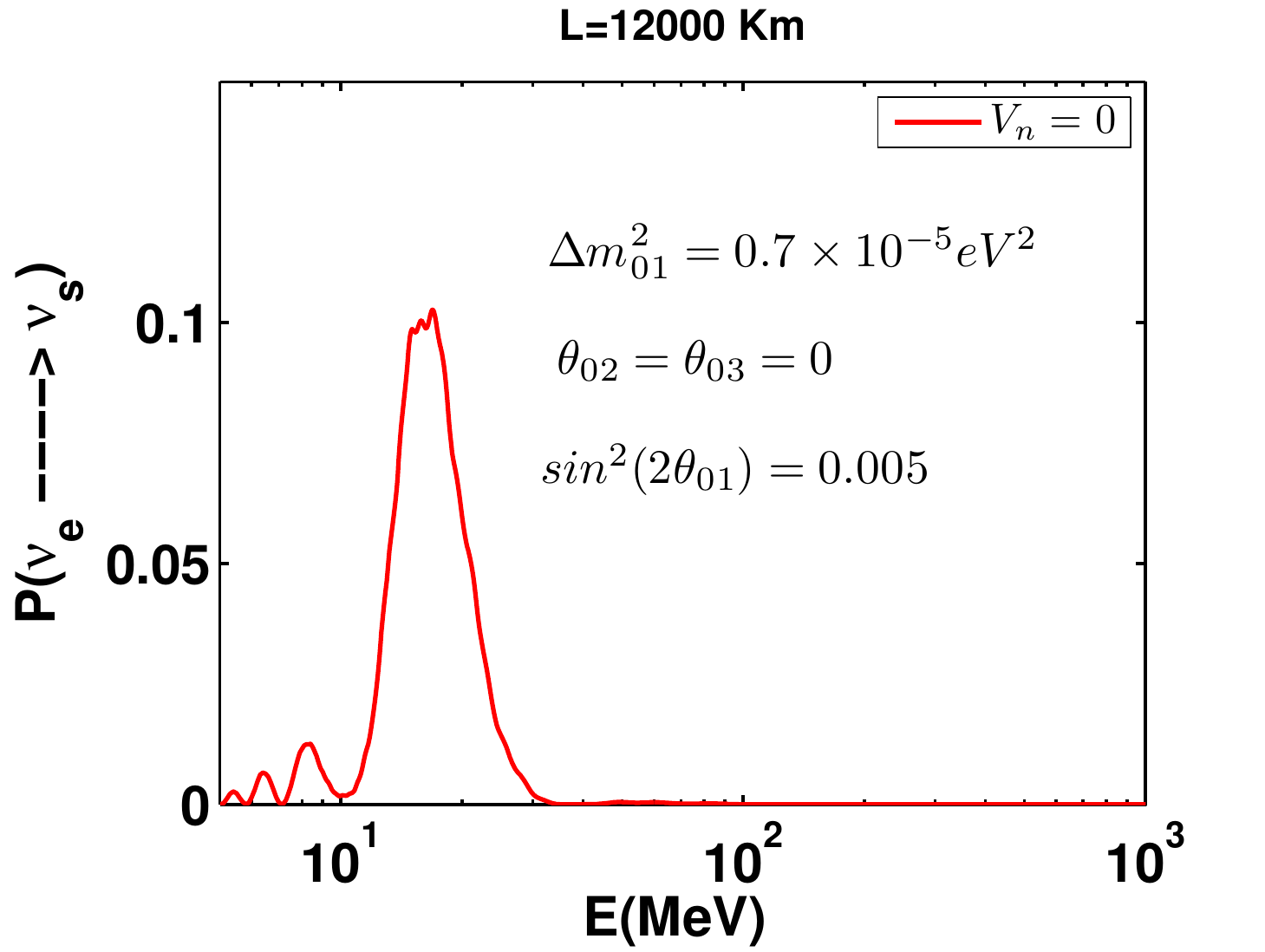}a)
\includegraphics[width=.3\textheight]{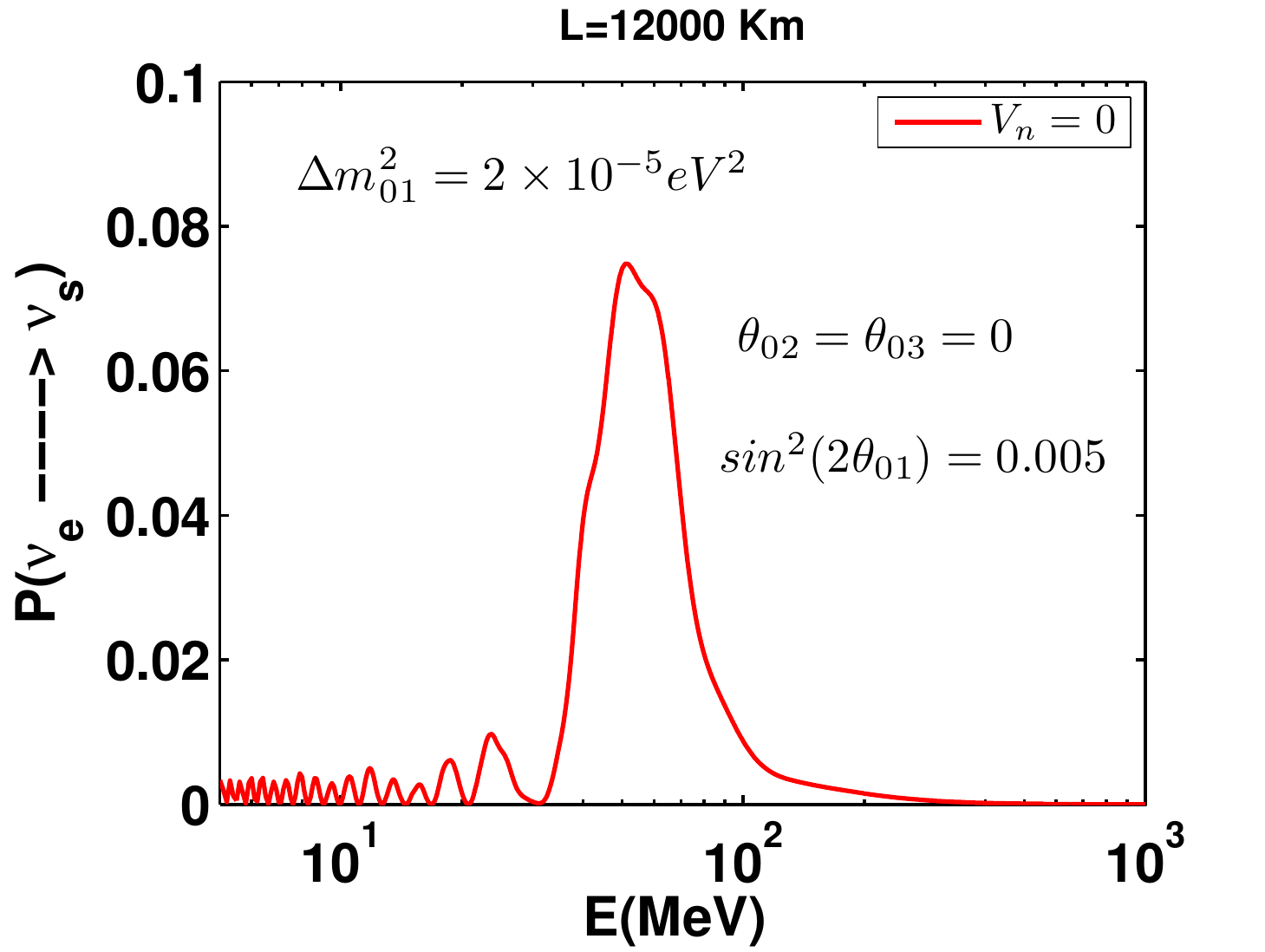}b)
\includegraphics[width=.3\textheight]{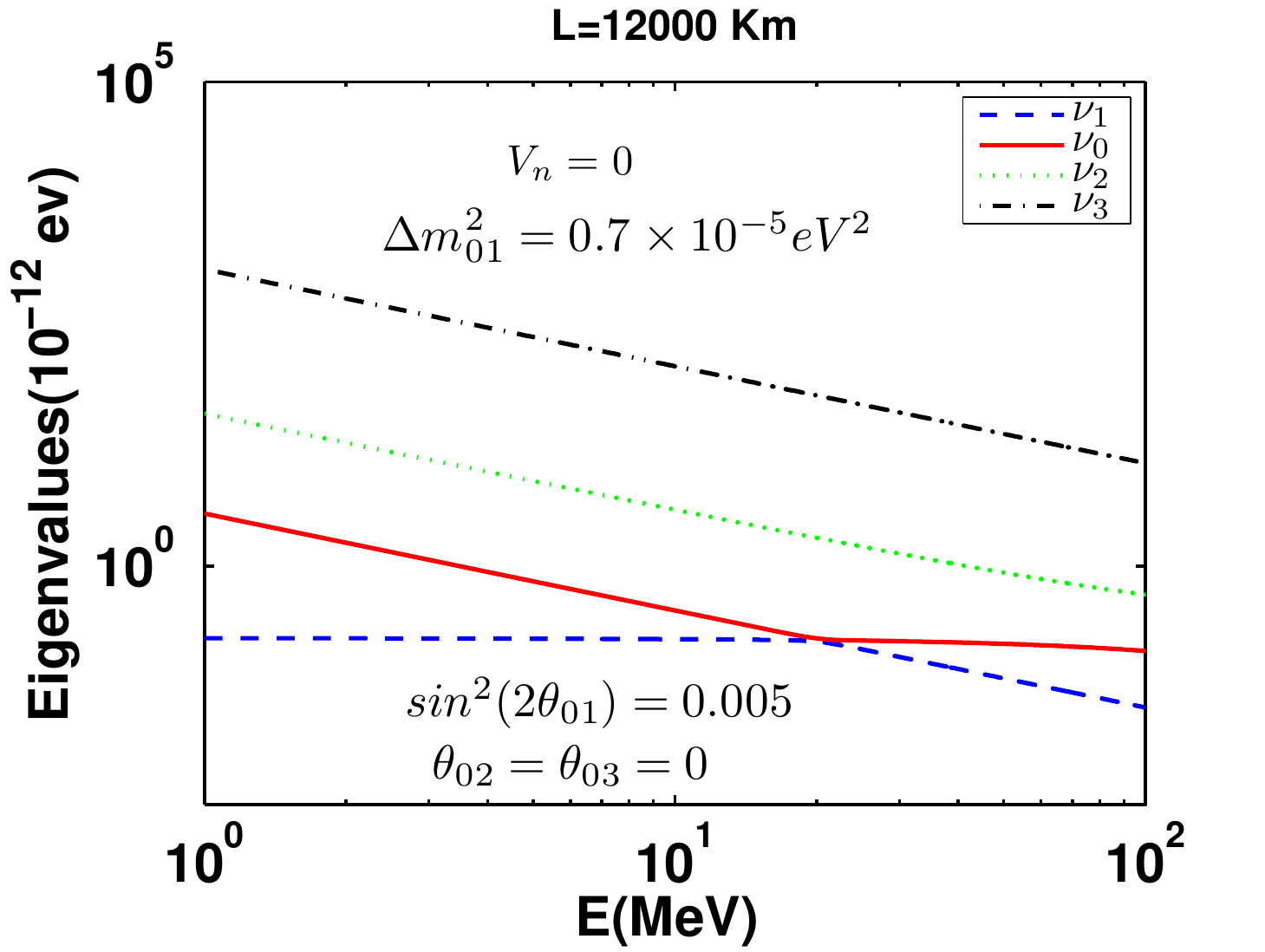}c)
\includegraphics[width=.3\textheight]{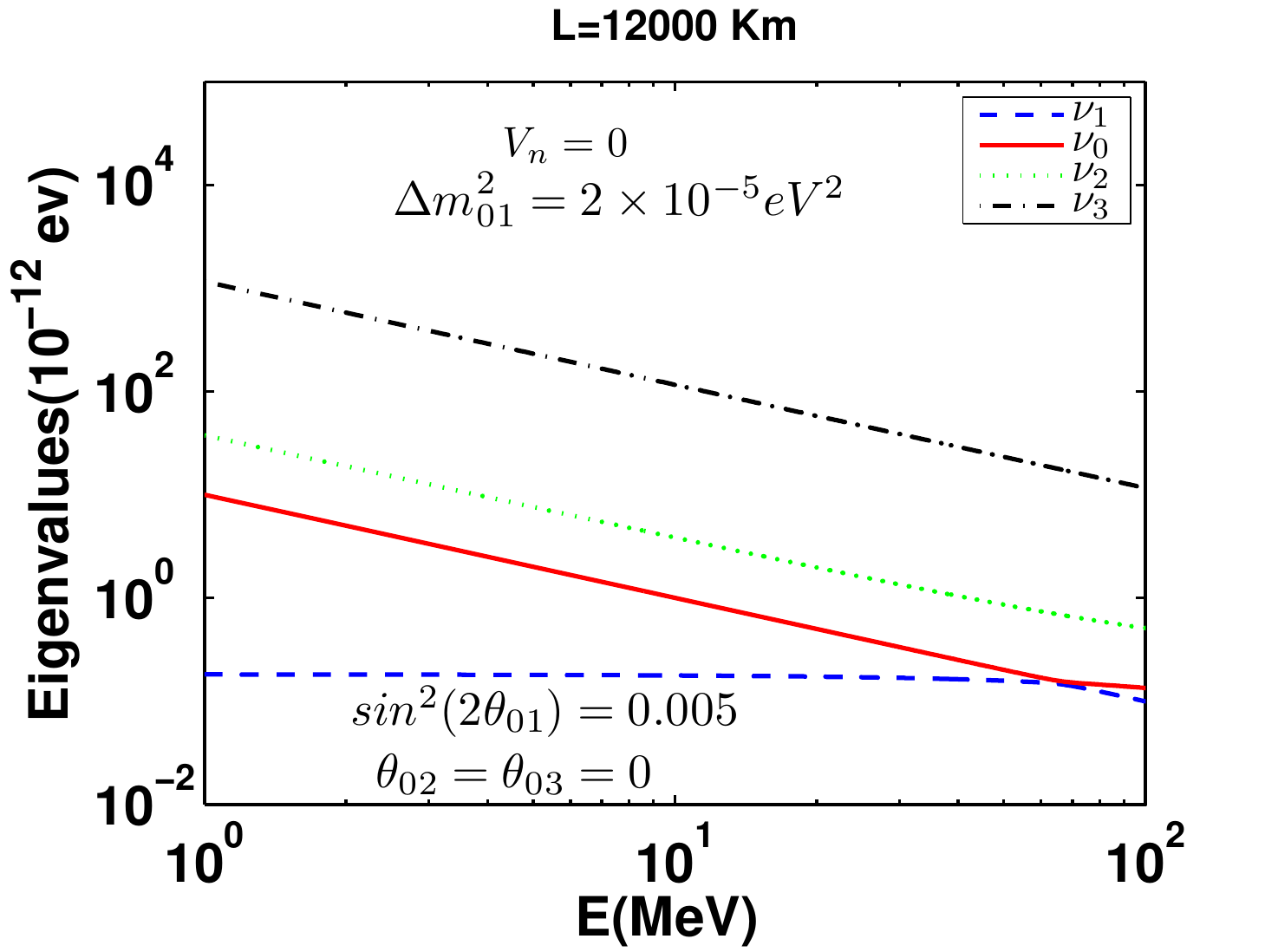}d)
\caption{(Color on line) Panels (a) and (b):
  $\nu_e\rightarrow\nu_s$ conversion probabilities versus neutrino  neutrino energy $E$ with
 $V_n=0$. Panels (c) and (d): Eigenvalues of
$\nu_0$, $\nu_1$, $\nu_2$ and $\nu_3$ versus neutrino energy $E$
with $V_n=0$.  Two different cases of the parameter ${\Delta
m}^2_{01}$ are considered at fixed  $L=12000$Km.}
 \label{eigen1}
  \end{figure}

\begin{figure}[htb]
\centering
\includegraphics[width=.3\textheight]{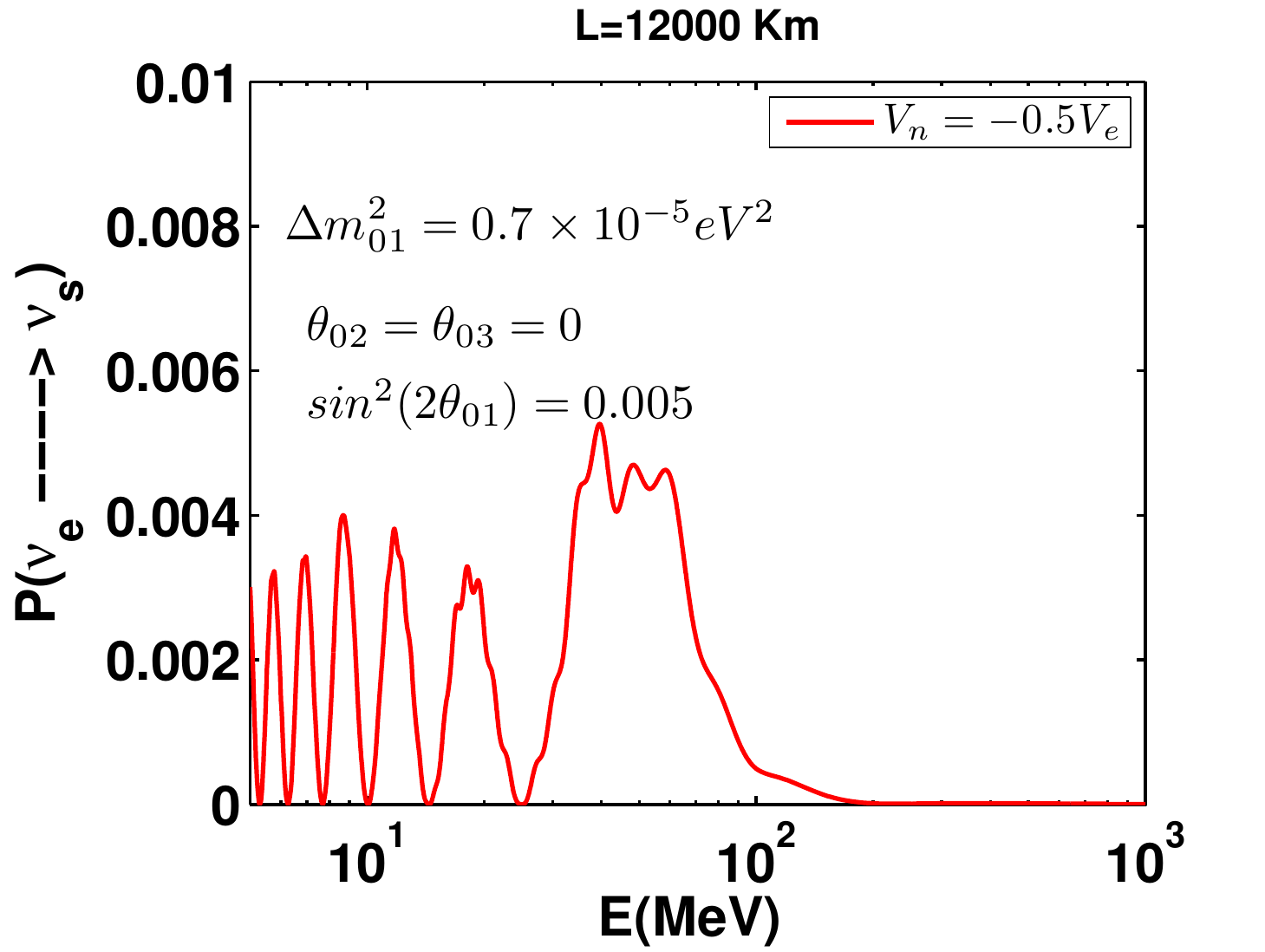}a)
\includegraphics[width=.3\textheight]{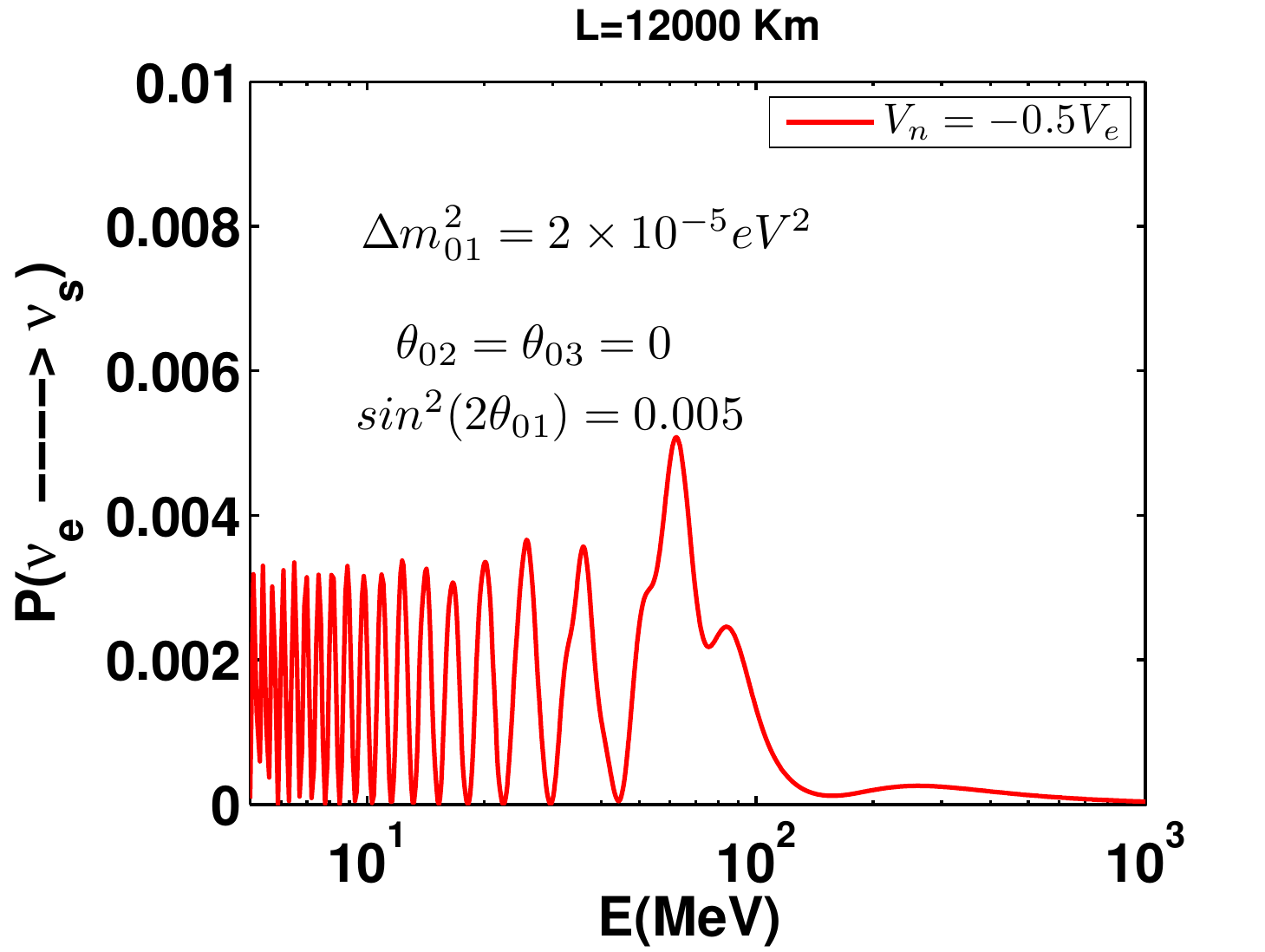}b)
\includegraphics[width=.3\textheight]{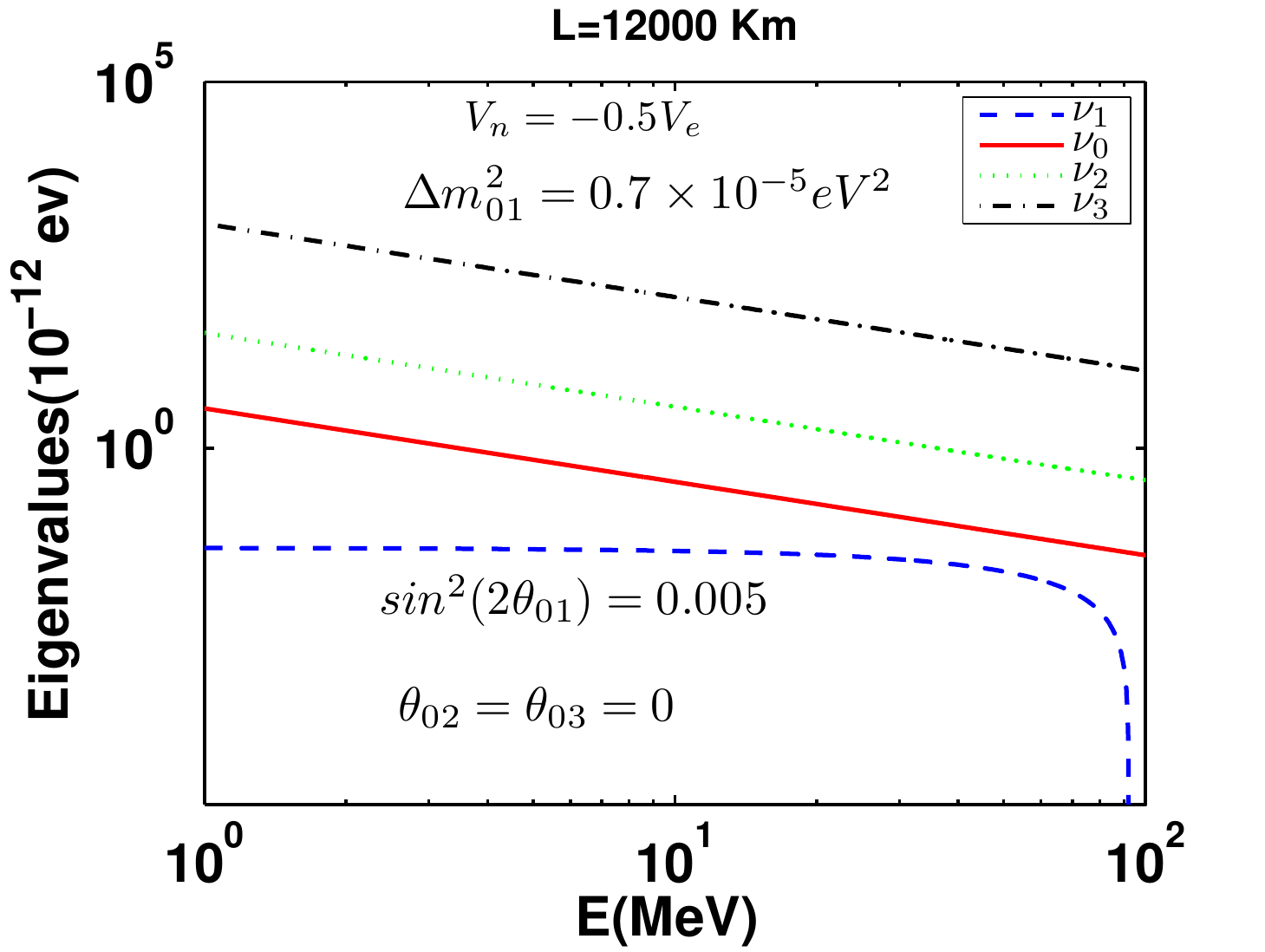}c)
\includegraphics[width=.3\textheight]{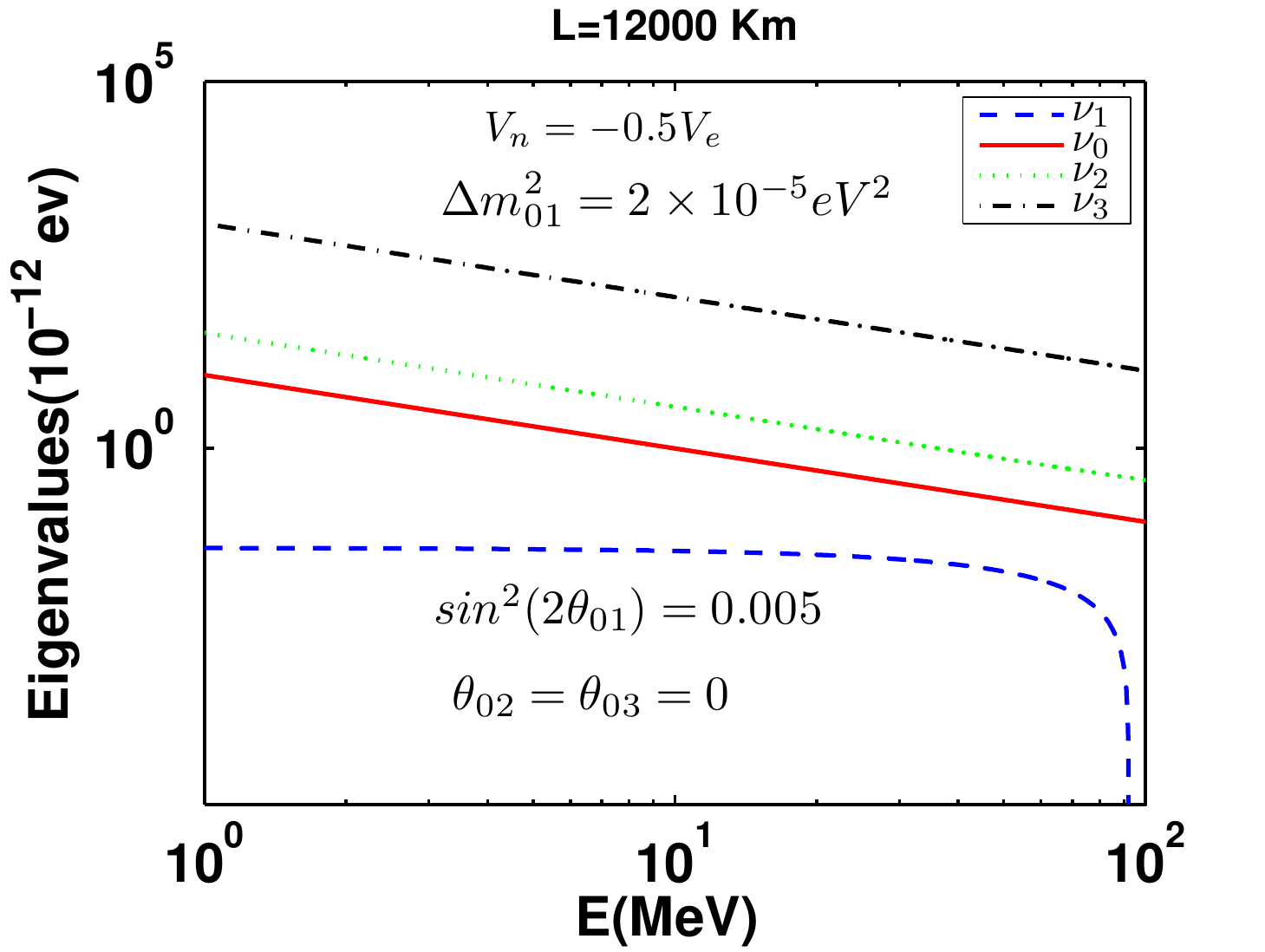}d)
\caption{(Color on line)  Panels (a) and (b):
  $\nu_e\rightarrow\nu_s$ conversion probabilities versus neutrino energy $E$ with
 $V_n=-0.5V_e$. Panels (c) and (d): Eigenvalues of
$\nu_0$, $\nu_1$, $\nu_2$ and $\nu_3$ versus neutrino energy $E$
with
 $V_n=-0.5V_e$.  Two different cases of the parameter ${\Delta m}^2_{01}$ are
considered. $L=12000$Km.}
 \label{eigen}
  \end{figure}

It is also interesting to study the effect of $V_n$ on the energy
levels of neutrinos and on the resonance conversion probability
$P(\nu_{e} \rightarrow \nu_s)$. In order to take into account
the Earth matter effect, it is convenient to compute the energy
levels of
\begin{equation}
H=U{H_0}U^\dagger +{\cal V} \label{hamil}
\end{equation}
taking a trajectory dependent
averaged potential $\bar V_e$ \cite{Liao}
 \begin{equation}
\bar V_e=\frac{1}{L}\int_0^L dx V_e(x)
  \label{averg}
\end{equation}
where $L$ is the length of the neutrino trajectory  in the Earth.
For baseline longer than 5000 km, ${\bar V}_e$ varies from
$1.36\times 10^{-13}$ eV to about $2.74\times 10^{-13}$ eV. For
neutrinos crossing  the core of  the Earth (approximately
$L=12000$ Km), ${\bar V}_e$ is found   to be  $2.74\times
10^{-13}$ eV.
 Figs.\hspace{2pt}\ref{eigen1} and \ref{eigen} show the
eigenvalues $E_0$, $E_1$,$E_2$ and $E_3$ corresponding to
neutrinos in the mass base $\nu_0$, $\nu_1$, $\nu_2$ and $\nu_3$
separately, as a function of the neutrino energy. We consider two
different cases of $\Delta { m}^2_{01}$. Also illustrated is the
conversion probability amplitude $\nu_e\rightarrow\nu_s$ versus
neutrino energy $E$ for the two individual  $V_n$ values. We note
that when a MSW resonance of flavor conversion takes place then
two of the energy levels of the neutrino  mass eigenstates $\nu_0$
and $\nu_1$ are getting close to each other. For $V_n=0$ the two
lines are crossing at a point with energy around 20 MeV for
$\Delta { m}^2_{01}=0.7 \times 10^{-5}$ eV$^{2}$ and 60 MeV for
$\Delta { m}^2_{01}=2 \times 10^{-5}$ eV$^{2}$, respectively.
 When $V_n=-0.5V_e$ (Fig.\hspace{2pt}\ref{eigen}) the two energy lines are drifting apart and the
resonance conversion probability has   significantly suppressed.
The absence of resonance is more clear as $\Delta { m}^2_{01}$
increases (Fig.\hspace{2pt}\ref{eigen}(b)) where   the resulting
neutrino oscillations are getting more rapid.

\begin{figure}[htb]
\centering
\includegraphics[width=.3\textheight]{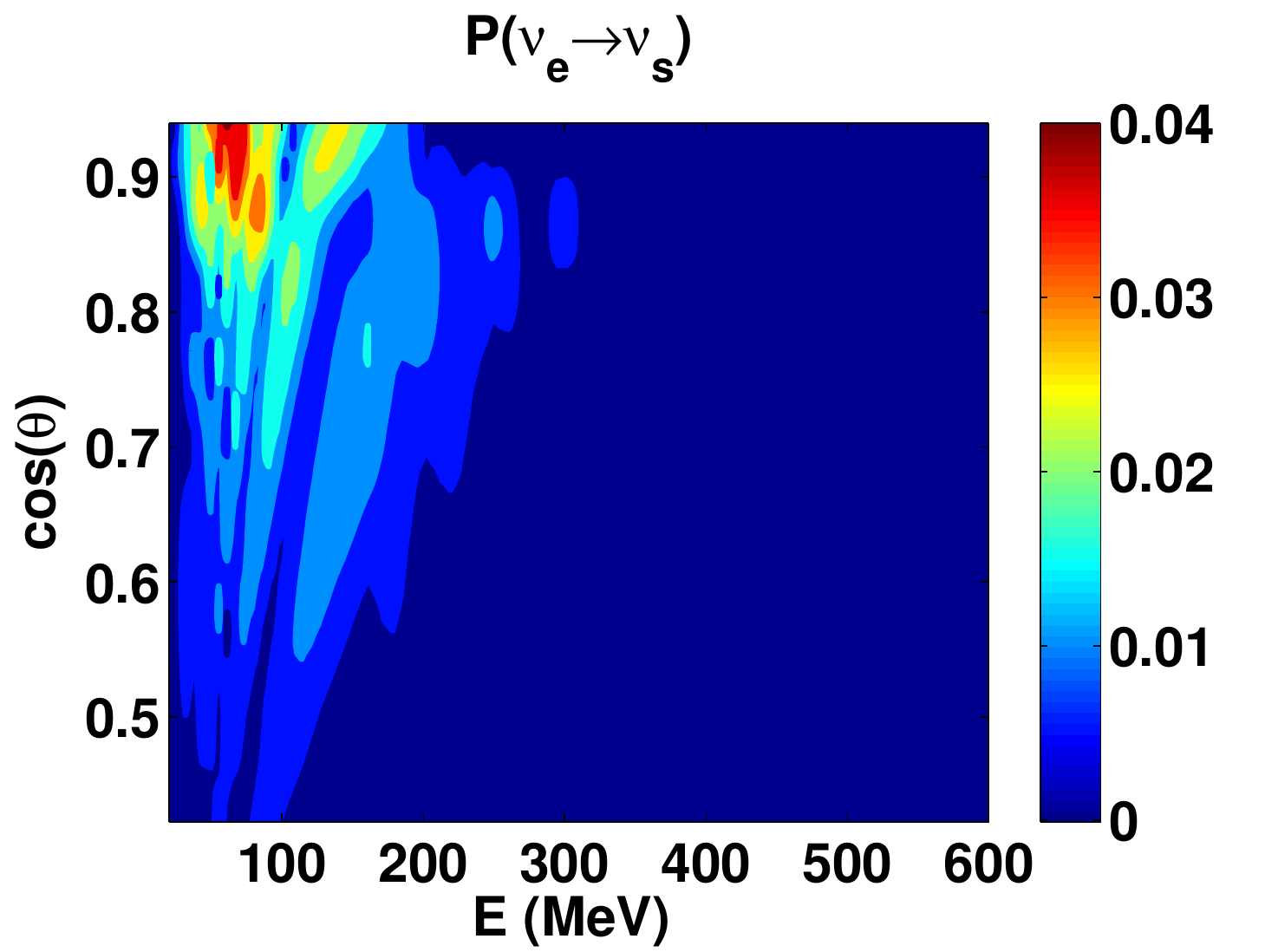}a)
\includegraphics[width=.3\textheight]{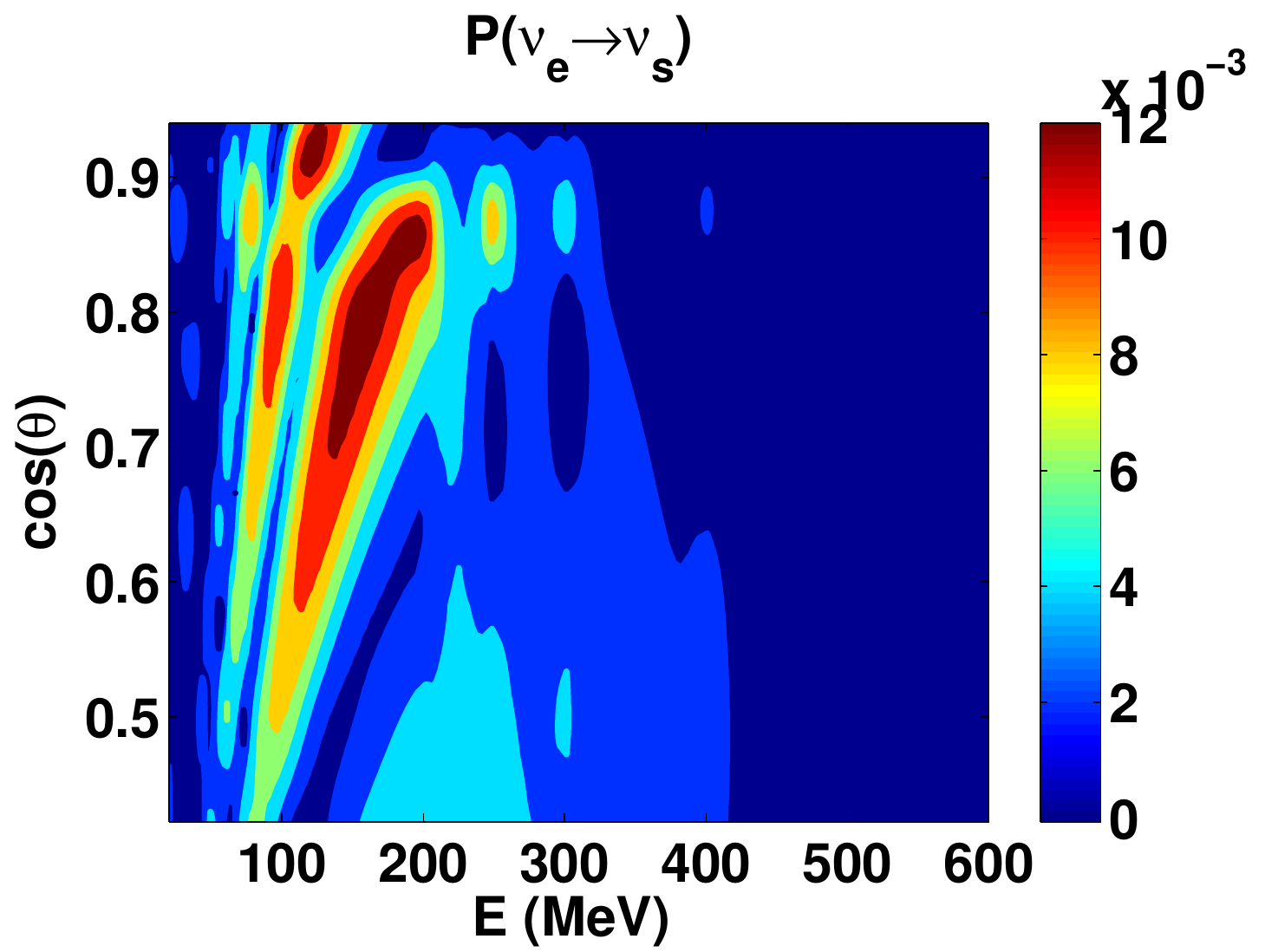}b)
\includegraphics[width=.3\textheight]{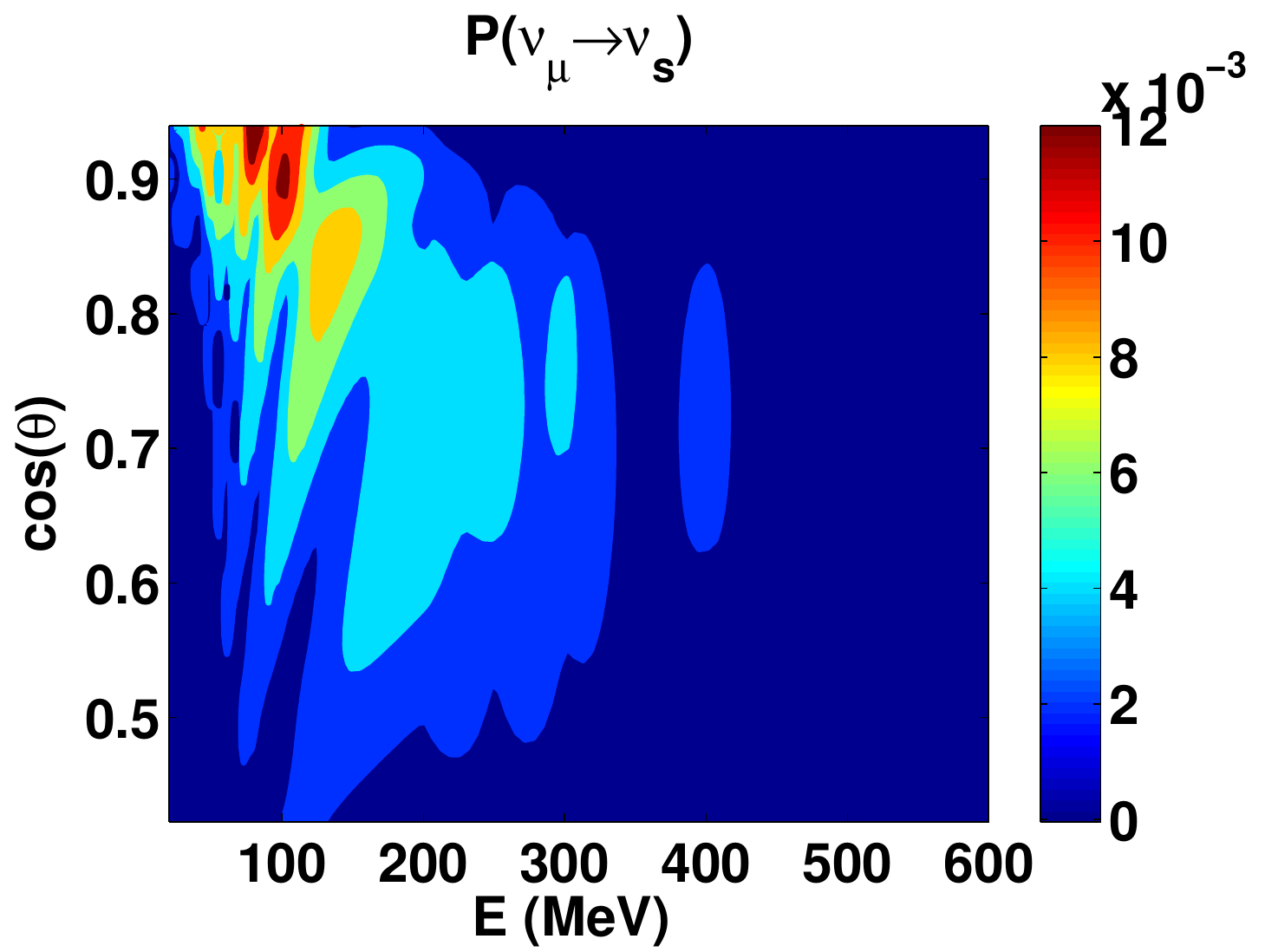}c)
\includegraphics[width=.3\textheight]{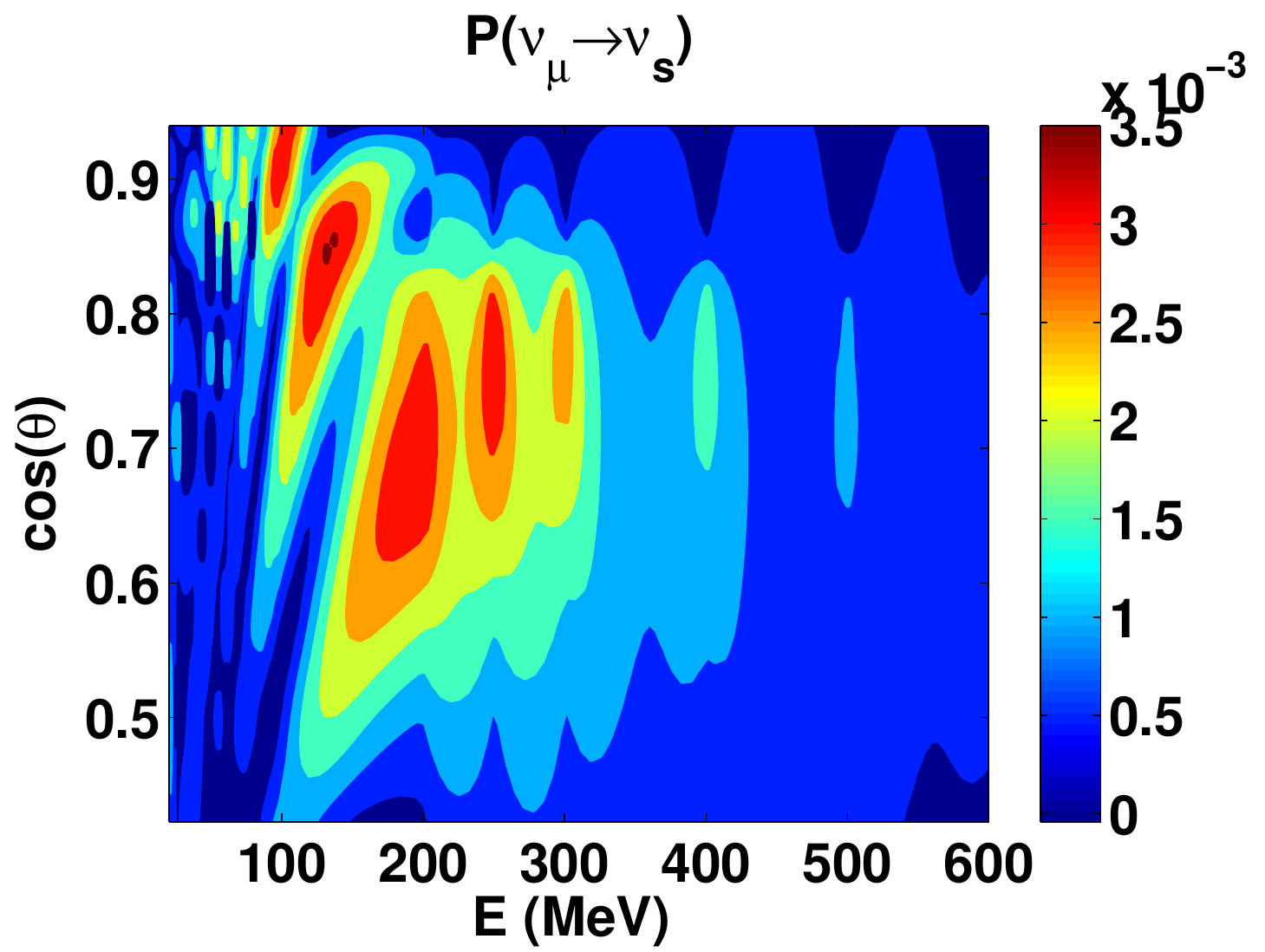}d)
\includegraphics[width=.3\textheight]{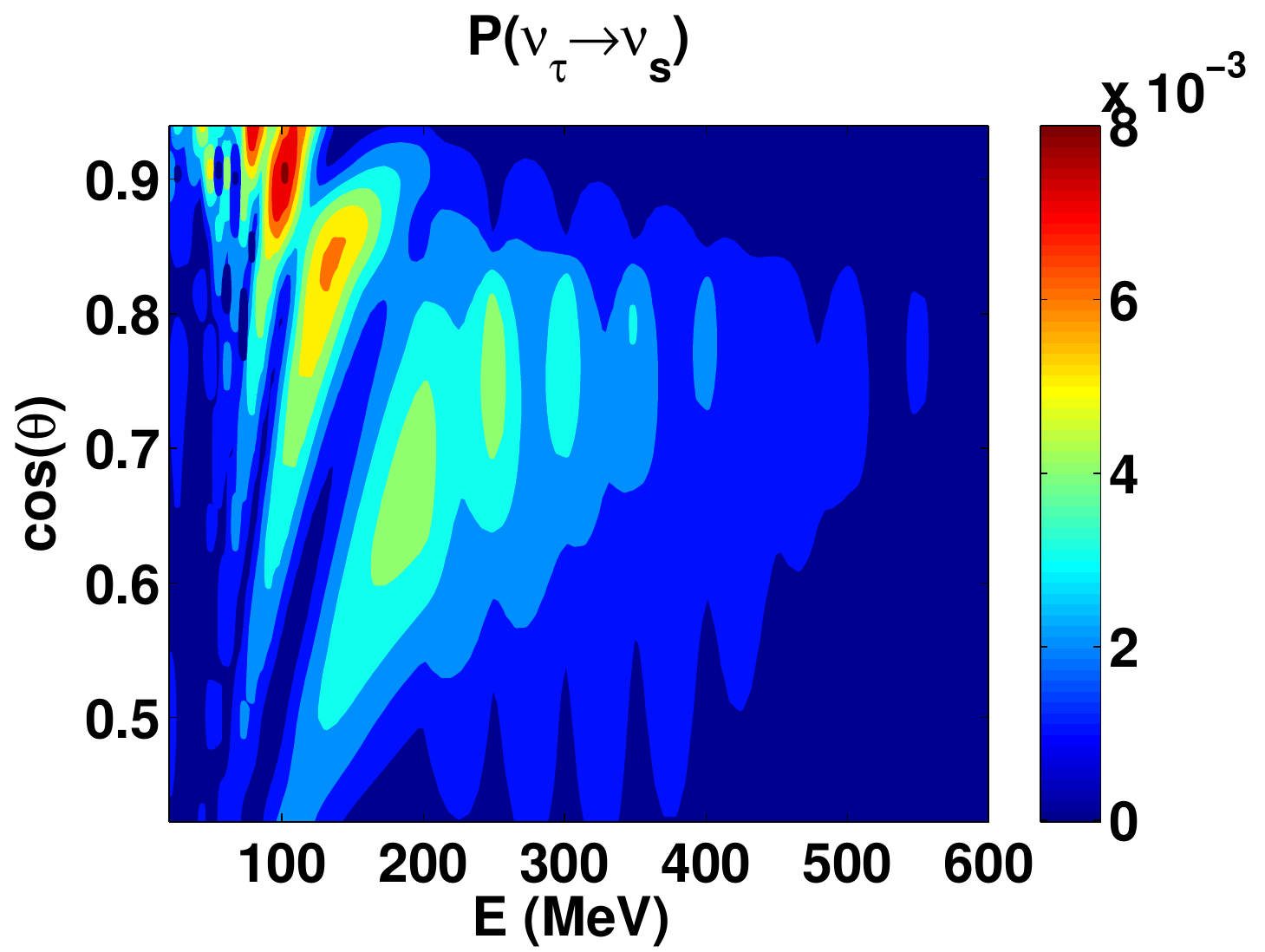}e)
\includegraphics[width=.3\textheight]{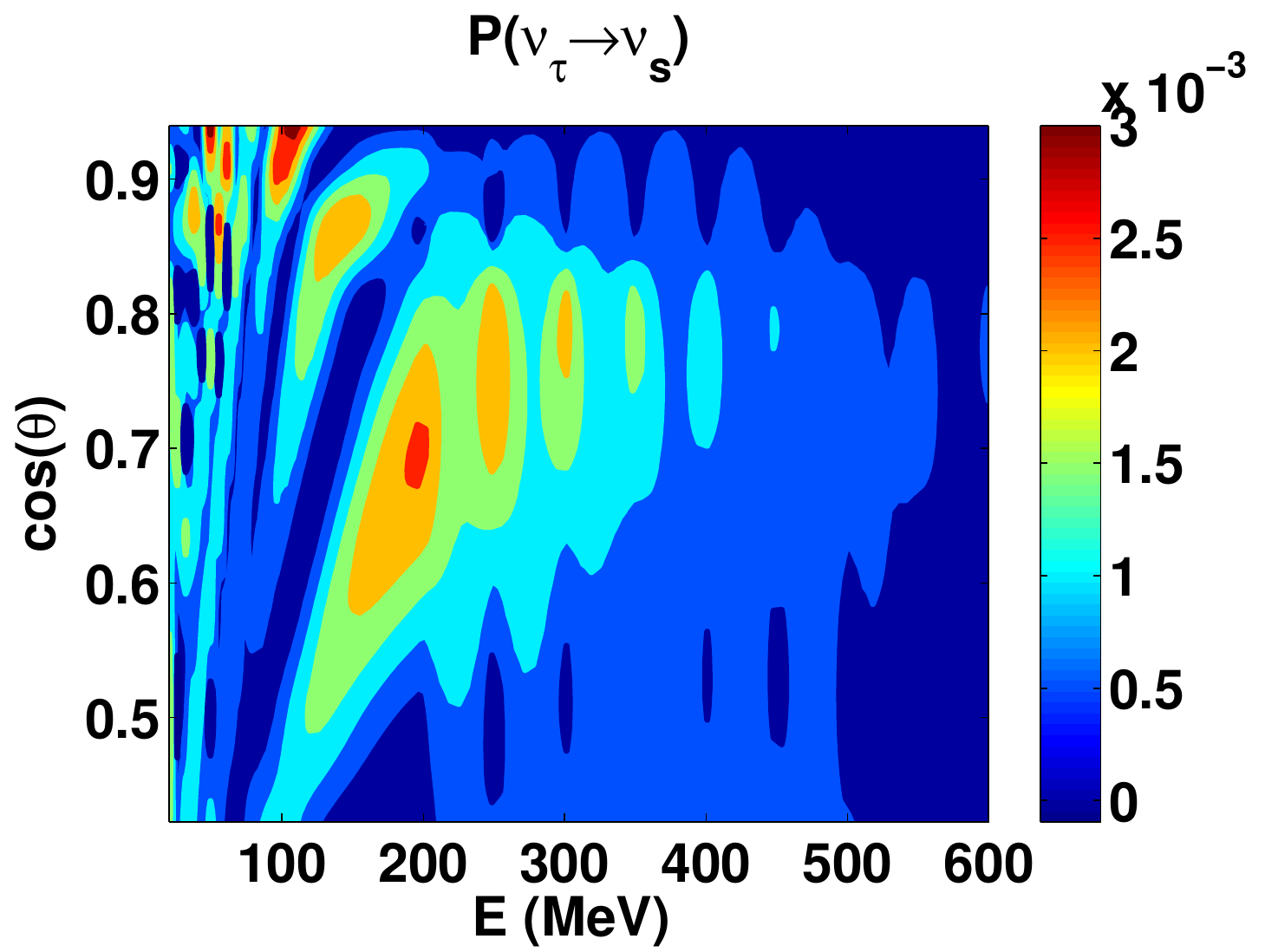}f)
\caption{(Color on line)
 Conversion probabilities as a function of nadir angle  $cos\hspace{2pt}\theta$ (y-axis) and
neutrino energy $E$ (x-axis).  The color represents the size of
the conversion probability. The  results are taken with
$V_n=-0.5V_e$. The sterile neutrino mixing parameters are:
$\theta_{01}=\theta_{03}=0$ and $\sin^2 2\theta_{02}=0.005$.
   $\Delta { m}^2_{01}=0.7 \times
10^{-5}$ eV$^{2}$ (left panels)  and   $\Delta { m}^2_{01}=1.5
\times10^{-5}$ eV$^{2}$ (right panels).}
 \label{lengthc}
  \end{figure}

Furthermore, Fig.\hspace{2pt}\ref{lengthc} shows contour plots of
$P(\nu_e,\nu_\mu,\nu_\tau\rightarrow\nu_s)$   as a function of
neutrino energy
  $E$ and  nadir angle
$cos\hspace{2pt}\theta$. The dark red   area   corresponds to
maximal conversion probability
$P(\nu_e,\nu_\mu,\nu_\tau\rightarrow\nu_s)$  and the dark blue to
very low one.  For $\Delta { m}^2_{01}=0.7 \times 10^{-5}$
eV$^{2}$ (left panels) the maximal conversion probability
(distinct red areas) occurs at lower energies (60-100) MeV and for
nadir angles $cos\hspace{2pt}\theta \simeq (0.85-0.95)$, that is,
for neutrinos travelling length approximately one Earth's
diameter.  As $\Delta { m}^2_{01}$ increases (right panels) a
broadening region of both $E$ and $cos\hspace{2pt}\theta$  is
indicated (the distinct red areas are slightly dissolved).
Moreover,  the oscillation pattern moves to higher $E$, around
(100-300)MeV, with oscillation amplitude being about three times
smaller.

It should be mentioned  that   all the above discussion has been
referred to the  disappearance probability
$P(\nu_{e,\mu,\tau}\rightarrow\nu_s)$ of active neutrinos from the
sterile one. For the appearance probability $P(\nu_s\rightarrow
\nu_{e,\mu,\tau})$ of active neutrinos from the sterile neutrino
implies that $P(\nu_s\rightarrow \nu_{e,\mu,\tau})=P(
\nu_{e,\mu,\tau}\rightarrow\nu_s)$, since the CP violating phase
in the mixing matrix $U$ is not taken into account.

\section{ CP-violation   effects in SLSN scenario}
\label{sec3a}

The four neutrino mixing matrix $U$ may also depends on
 three physical Dirac phases, the standard phase $\eta_{13}$, and
 the two nonstandard ones $\eta_{02}$ and $\eta_{03}$, involved in the
 SLSN scenario.  Taking into account the CP-violation phases in Eq.~(\ref{H0c}), U is  written
 \begin{eqnarray}
U=R(\theta_{23}) \widetilde{R}(\theta_{13}) R(\theta_{12})
\widetilde{R}(\theta_{02}) R(\theta_{01})
\widetilde{R}(\theta_{03}) \label{H0cs},
\end{eqnarray}
with
\begin{eqnarray} \widetilde{R}(\theta_{02})=\begin{pmatrix} c_{02} & 0&
\widetilde{s}_{02}  & 0 \cr
 0 & 1 & 0 & 0 \cr
 -\widetilde{s}_{02}  & 0 & c_{02} & 0  \cr
 0 & 0 & 0 & 1 \cr \end{pmatrix}, \label{R02s}
\end{eqnarray}
\begin{eqnarray}
 \widetilde{R}(\theta_{13})=\begin{pmatrix} 1 &
0 & 0 & 0 \cr 0 & c_{13} & 0 & \widetilde{s}_{13} \cr
 0 & 0 & 1 & 0 \cr 0 & -\widetilde{s}_{13} & 0 & c_{13} \cr \end{pmatrix},\label{R13s}
 \end{eqnarray}
\begin{eqnarray}
\widetilde{R}(\theta_{03})=\begin{pmatrix} c_{03} & 0& 0 &
\widetilde{s}_{03} \cr
 0 & 1 & 0 & 0 \cr
 0 & 0 & 1 & 0  \cr
 -\widetilde{s}_{03} & 0 & 0 & c_{03} \cr \end{pmatrix}, \label{R03s}
 \end{eqnarray}
where
 $\widetilde{s}_{ij}=s_{ij}e^{-\imath \eta_{ij}}$, with $s_{ij}=\sin\theta_{ij}$ and
 $c_{ij}=\cos\theta_{ij}$

\begin{figure}[htb]
\centering
\includegraphics[width=0.3\textheight]{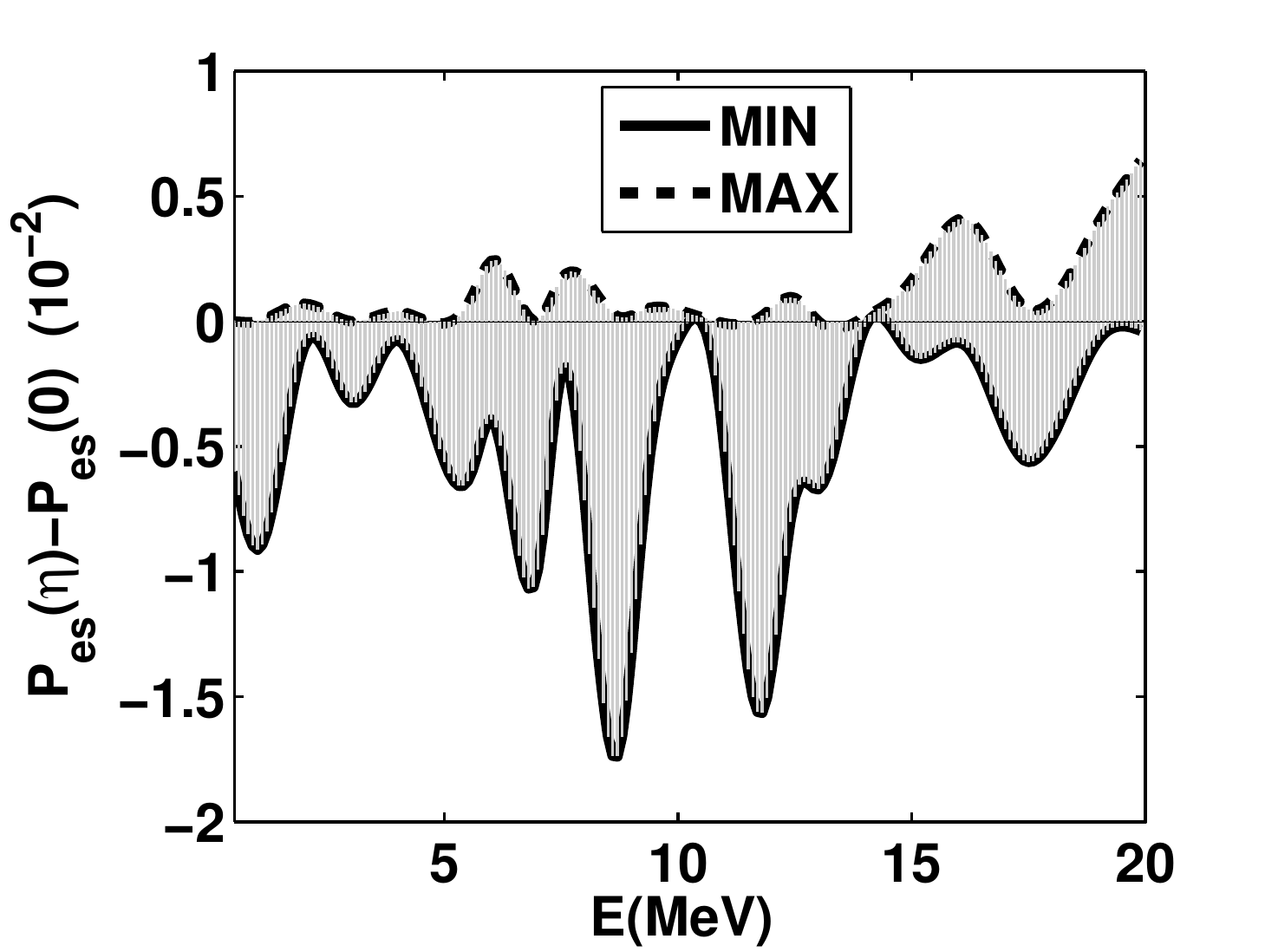}a)
\includegraphics[width=0.3\textheight]{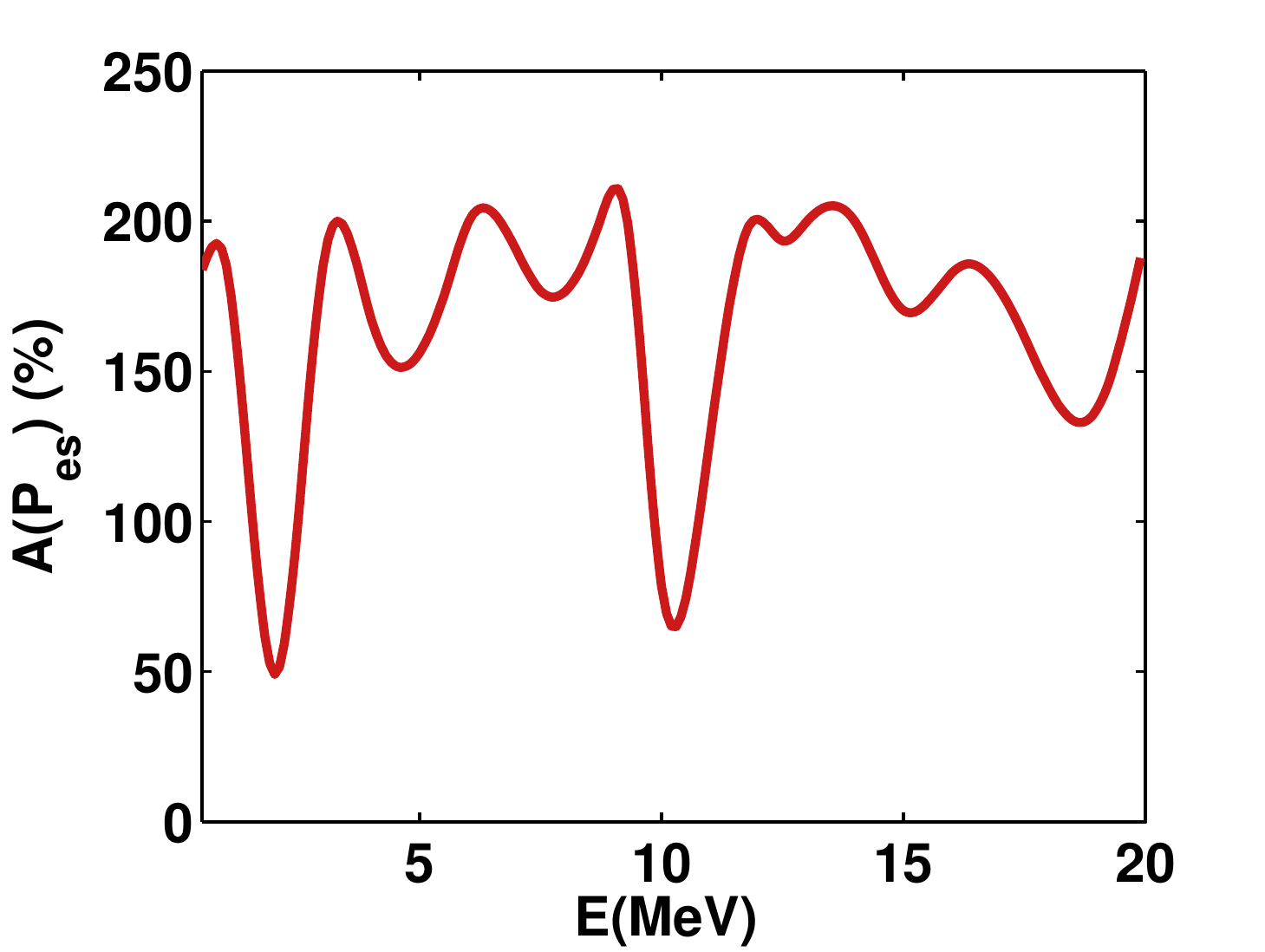}b)
\caption{(Color on line) Energy spectra of
$P_{es}({\eta})-P_{es}(0)$ and $A(P_{es})$.
  The mass and mixing
 parameters relevant with SLSN are set to  $\Delta {
m}^2_{01}=0.7\times 10^{-5}$ eV$^2$ and $\sin^2
2\theta_{01}=\sin^2 2\theta_{02}=\sin^22\theta_{03}=0.005$.
$L=12000$Km.}
 \label{cpes}
  \end{figure}

\begin{figure}[htb]
\centering
\includegraphics[width=0.3\textheight]{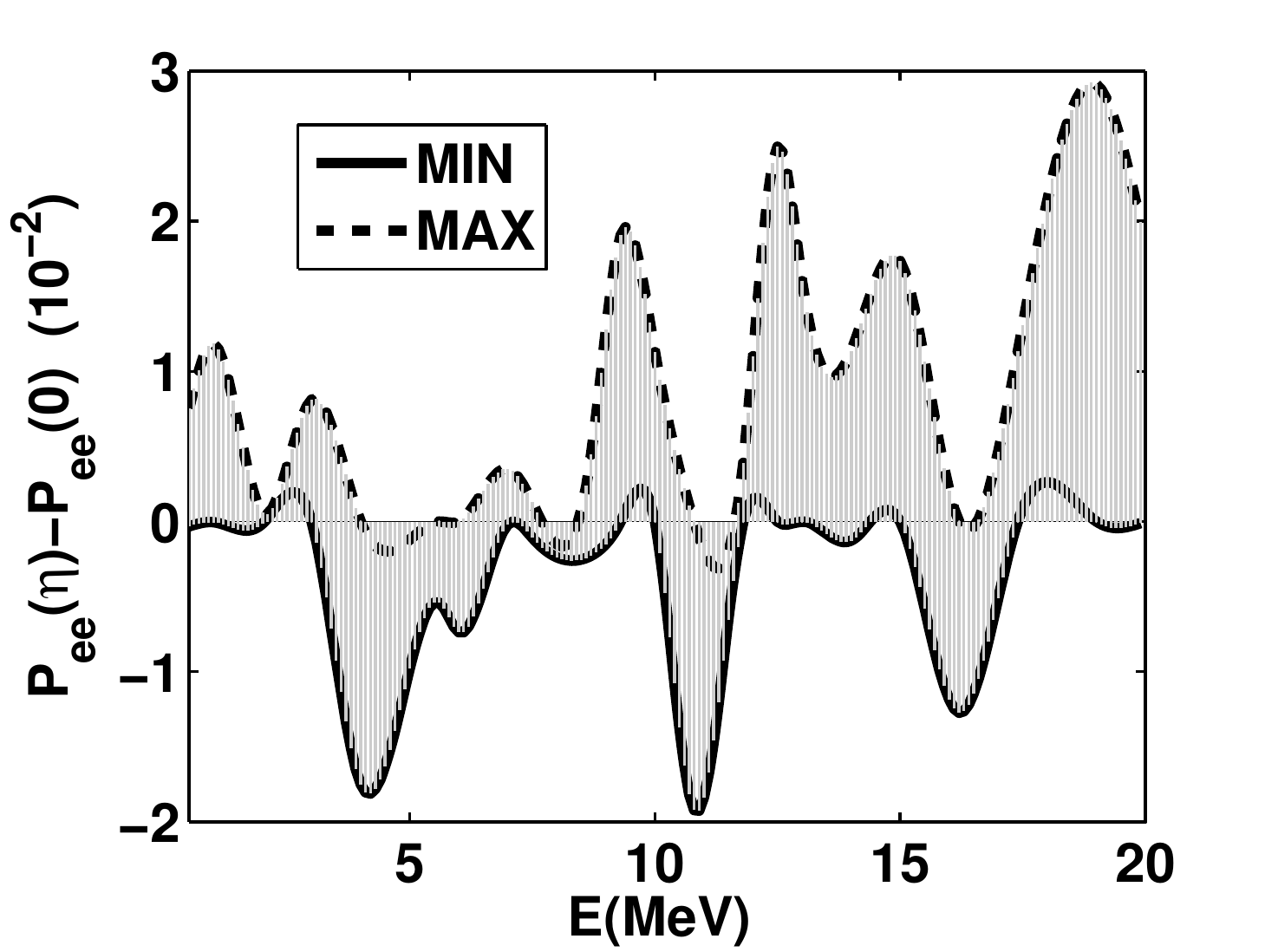}a)
\includegraphics[width=0.3\textheight]{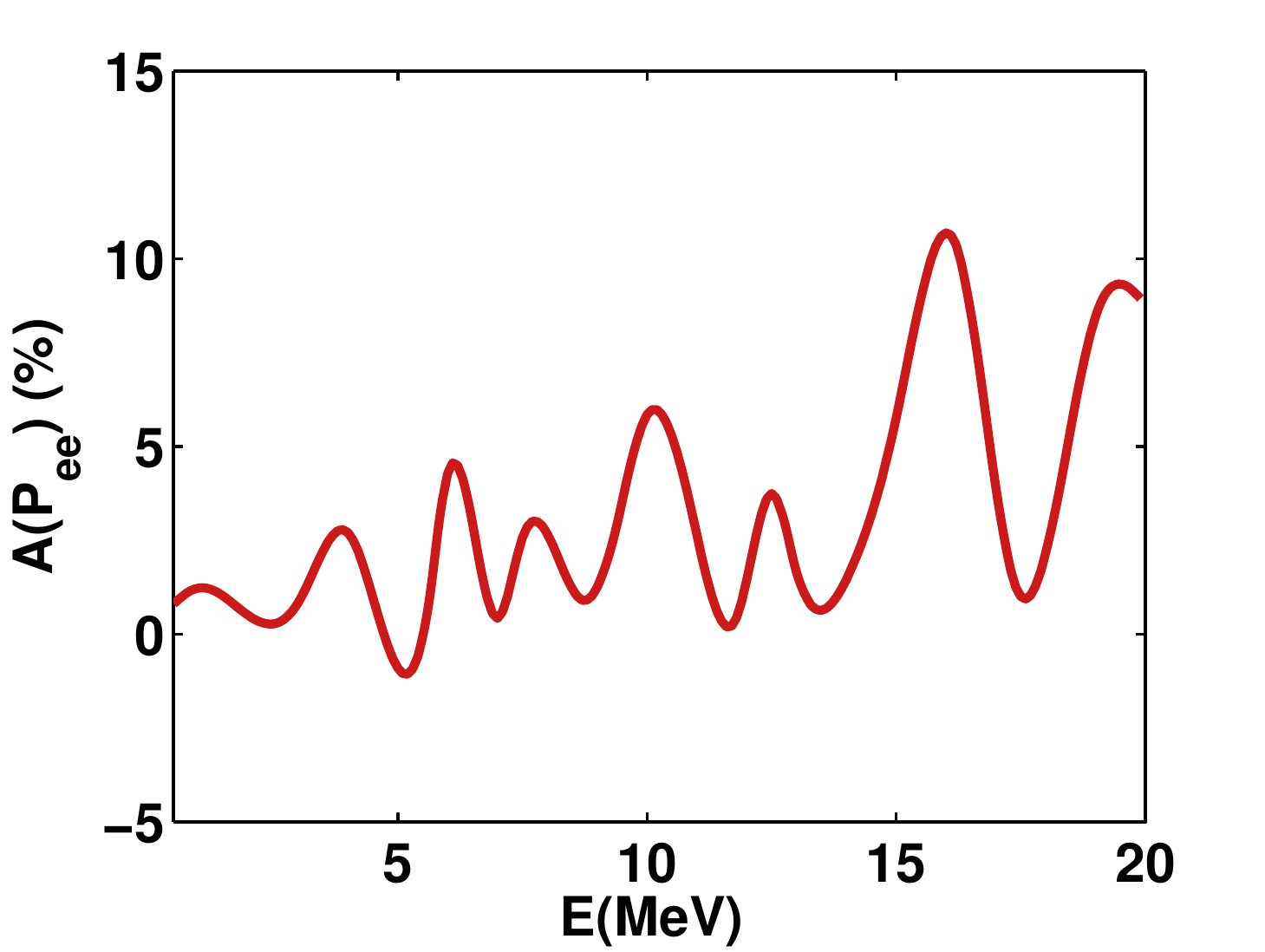}b)
\caption{(Color on line) Same as  Fig.~\protect\ref{cpes}, but for
electron survival probability. }
 \label{cpee}
  \end{figure}

 Figure \ref{cpes}(a)  displays  the energy
 spectra of the electron to super light sterile transition  probability  $P_{es}$ for all possible values   of the
 CP-violating phases, in the range $[-\pi, \pi]$, with respect to
 the case $\eta_{03}=\eta_{02}=\eta_{13}=0$. The results are given
 in the energy region of measurable solar neutrinos, from 0.5 MeV to 20 MeV.
 The shadowed region
 is generated by the full parameter space of the three
 CP-violating phases.  The boundary curves stand for the maximal and minimal values of the
 difference $P_{es}({\eta})-P_{es}(0)$. Moreover,
 these maximal and minimal
 values can be used to build a CP-violating asymmetry $A(P_{es})$
 (see  Ref.~\cite{long1,long2})
\begin{equation}
\label{asym} A(P_{es})=2\times
\frac{MAX(P_{es})-MIN(P_{es})}{MAX(P_{es})+MIN(P_{es})}
\end{equation}
displayed in Fig.~\ref{cpes}(b) as a function of solar neutrino
energy.  In Fig.~\ref{cpee}(a) and (b) we have also shown  the
curves corresponding to electron neutrino survival probability
$P_{ee}({\eta})-P_{ee}(0)$ and $A(P_{ee})$, respectively. As it is
seen,   the variation induced by the three CP-violating phases is
less than 10\% for the electron survival probability and can reach
the level of 200\% for the electron-to-sterile transition
probability. This significant departure from zero in the asymmetry
could   be interpreted as a manifestation of leptonic CP
violation. Future neutrino facilities could be a powerful tool to
accurately assess the values of the elements of the mixing matrix
$|U_{a0}|$ for $a=e,\mu,\tau$. In this case, it might be possible
to observe the effects of the CP-violating phases in future solar
neutrino experiments.

\section {Super light sterile neutrino oscillation  searches at nuclear reactors}
\label{sec4}

Nuclear reactors are intense, isotropic sources of $\bar\nu_e$
produced by $\beta$-decay of fission fragments (i.e. U and Pu),
into more stable nuclei: $^A_ZX\rightarrow
^A_{Z-1}Y+e^-+\bar\nu_e$. The $\bar\nu_e$ energy is below 10 MeV,
with an average value of $\sim 3$ MeV.

The neutrino oscillation search at a reactor
(see Ref.~\cite{Vogel} and references therein)
 is
 based on a disappearance measurement using the  detection
process $\bar\nu_e+p\rightarrow e^+ + n$. Although the energy of
the reactor  neutrinos is of the order of a few MeV and the
interaction cross section between matter and reactor antineutrinos
is very tiny ($~10^{-44}$ cm$^2$), the huge emitted flux ($2\times
\hspace{1pt}10^{20}$ antineutrinos/second from a 1GW reactor)
allows us to detect their signal. At such energies matter effects
 on the oscillation probability are negligible: $V_{eff}\sim
  G_F N_e \sim G_F N_n\ll \Delta m_{01}^2/E <\Delta
  m_{21}^2/E\ll |\Delta m_{31}^2/E|$. The  oscillation  probability is given simply by
  \begin{equation}
  \label{PeeBB}
  P(\bar{\nu}_e \to \bar{\nu}_e)
=\left|M_0e^{i \delta_{01}}+M_1 +M_2e^{i \delta_{21}}+M_3e^{i
\delta_{31}}\right|^2
\end{equation}
where $\delta_{ij} =\Delta m_{ij}^2L/2E$ and
\begin{eqnarray} \label{M0-3}M_0&=&|c_{03} (-c_{01} s_{12} c_{13}s_{02} -
c_{13}s_{01} c_{12})- e^{i\delta_D}s_{03} s_{13}|^2, \cr
M_1&=&(c_{13}c_{12}c_{01}-c_{13}s_{12}s_{01}s_{02})^2 \cr M_2&=&(
s_{12}c_{13}c_{02})^2 \cr M_3 &=& |s_{03}c_{13} (-c_{12}s_{01}
-c_{01}s_{02} s_{12}) + e^{i\delta_D}c_{03} s_{13}|^2.
\end{eqnarray}
$\delta_D$ being the Dirac CP-violating phase and
$c_{ij}=\cos\theta_{ij}$,   $s_{ij}=\sin\theta_{ij}$,
$c_{oi}=\cos\theta_{0i}$,  $s_{oi}=\sin\theta_{0i}$ \hspace{1pt}
$i,j=1,2,3$. In the absence of mixing with the sterile neutrinos,
we recover the standard formula:
\begin{equation} \label{os} P(\bar{\nu}_e \to \bar{\nu}_e)
=\left|\cos^2\theta_{13}\cos^2\theta_{12}+\cos^2\theta_{13}
\sin^2\theta_{12}e^{i \delta_{21}}+\sin^2 \theta_{13}e^{i
\delta_{31}}\right|^2\end{equation}
  \begin{figure}[htb]
\centering
\includegraphics[width=.3\textheight]{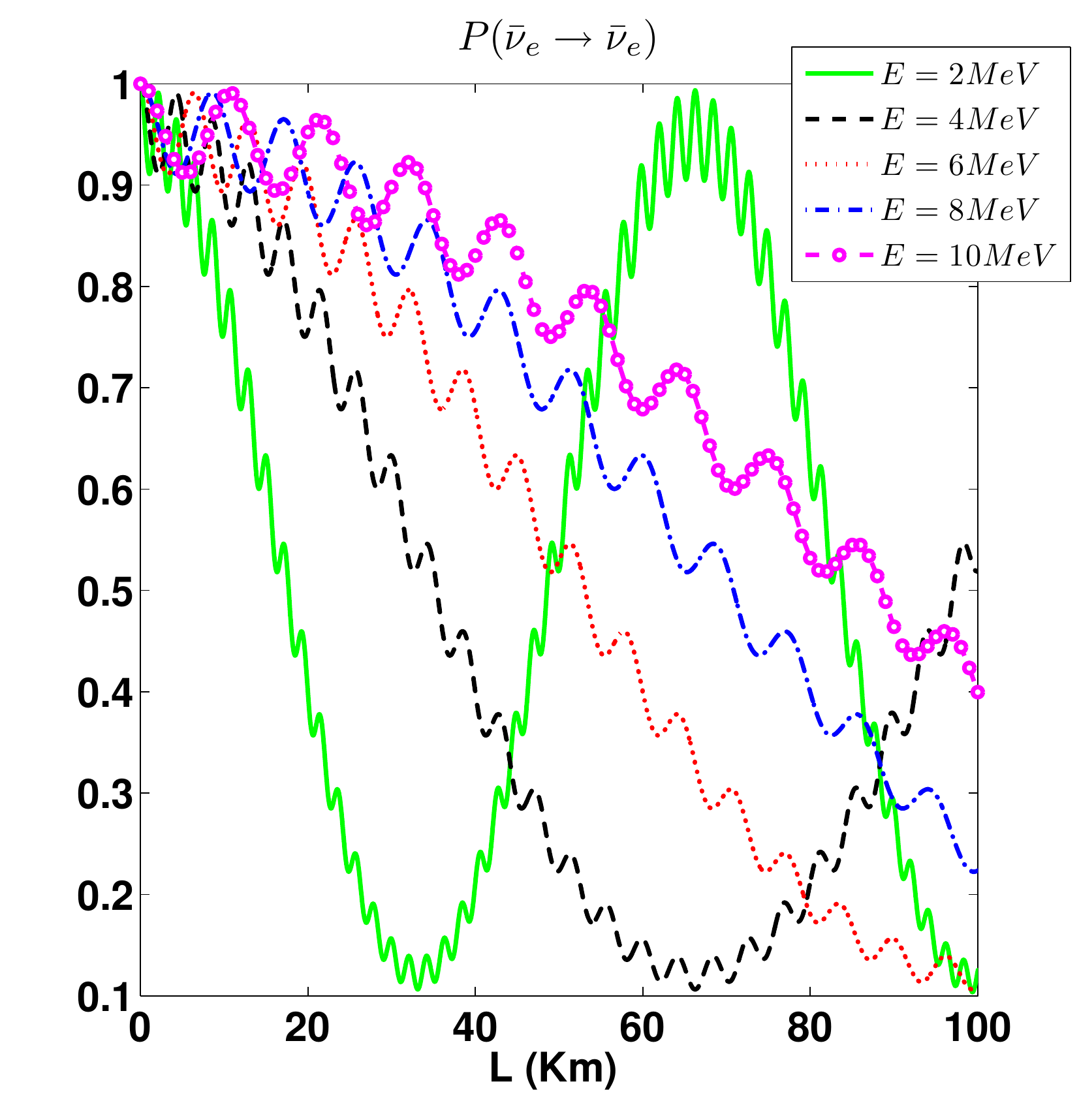}
\caption{(Color on line).   Oscillation pattern in $P(\bar{\nu}_e
\to \bar{\nu}_e)$  probability  as a function of baseline L, for
various neutrino energies. The  parameters used are : $\Delta {
m}^2_{01}=2 \times 10^{-5}$ eV$^{2}$ and $\sin^2
2\theta_{0i}=0.005$, \hspace{2pt} i=1,2,3. }
 \label{ster}
  \end{figure}
  The oscillation probability $P(\bar{\nu}_e \to
\bar{\nu}_e)$     as a function of baseline L for various neutrino
energies $E$ is depicted in  Figure \ref{ster}. The oscillation
probability $P(\bar{\nu}_e \to\bar{\nu}_e)$ beats because the
frequency of the waves in (\ref{PeeBB}) with $|\delta m_{13}^2|$
differs from the frequencies associated with $|\delta m_{12}^2|$
and $|\delta m_{01}^2|$, which are smaller by a factor of $ 30$
and $ 100$ respectively compared to $|\delta m_{13}^2|$ scale.
The superposed waves can be decomposed  into the beating low
frequency wave and the high frequency wiggles within the beat. The
frequency of wiggles is higher for smaller neutrino energies while
the ratio of the
 period $P$ of the  wiggles to  neutrino energy $E$, is $P/E\simeq
1$. We  also note that at low $E$ the oscillation wiggles grow
smaller with higher baseline $L$. Moreover,    the lower $E$ is
the more rapid the wiggles are in $L$, while as the neutrino
energy $E$ increases  wiggling formation   is diminishing.

\begin{figure}[htb]
\centering
\includegraphics[width=.4\textheight]{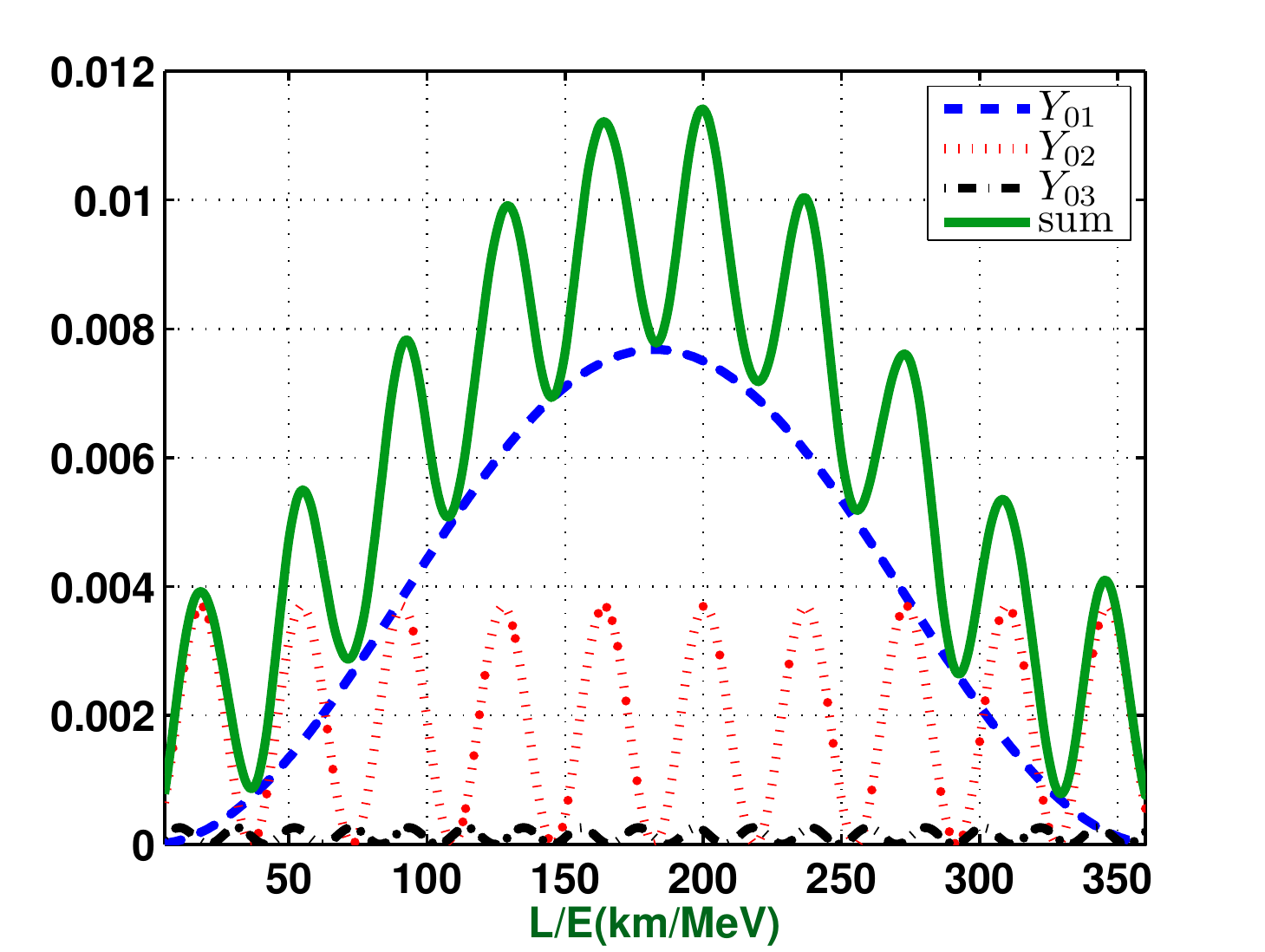}a)
\caption{(Color on line). Oscillatory modes
$Y_{0i}=4M_0M_i\sin^2\Delta_{0i}$ as function of   $L/E$. The mass
splitting parameter
 is taken $\Delta { m}^2_{01}=0.7 \times
10^{-5}$ eV$^{2}$. The sterile mixing parameters are: $ \sin^2
2\theta_{0i}=0.005, \hspace{2pt} i=1,2,3$. The solid (green) line
depicts the summation $Y_{01}+Y_{02}+Y_{03}$.
}
 \label{oscf}
  \end{figure}

Let us now discuss the effects of  mixing parameters by the
sterile neutrino.
 According to the general expression in Eq.
(\ref{PeeBB}), the electron antineutrino
 survival probability   $P(\bar{\nu}_e \to
\bar{\nu}_e)$ reduces into the form
\begin{equation} \label{os-reduced}
P(\bar{\nu}_e \to \bar{\nu}_e)=
1-4\sum_{i=1}^3M_0M_i\sin^2\Delta_{0i}-4\sum_{i>j=1}^3M_iM_j\sin^2\Delta_{ij}
\end{equation}
where $\Delta_{ij}=\Delta m_{ij}^2L/4E$.  Eq.~(\ref{os-reduced})
includes  seven oscillatory modes written as
\begin{eqnarray}
\label{oscn}P(\bar{\nu}_e \to
\bar{\nu}_e)&=&1-(Y_{01}+Y_{02}+Y_{03}+Y_{21}+Y_{31}+Y_{32})
\end{eqnarray}
where
\begin{equation}
\label{osc01}
Y_{01}=0.008sin^2\Big(\frac{\pi}{\Lambda_{01}}\frac{L}{E}\Big),\quad
\Lambda_{01}\simeq354 km/MeV\hspace{3pt}
\end{equation}
\begin{equation}
\label{osc02}
Y_{02}=0.004sin^2\Big(\frac{\pi}{\Lambda_{02}}\frac{L }{E}\Big),
\quad \Lambda_{02}\simeq 36.5 km/MeV \hspace{3pt}
\end{equation}
\begin{equation}
\label{osc03}
Y_{03}=0.0002sin^2\Big(\frac{\pi}{\Lambda_{03}}\frac{L }{E}\Big),
\quad  \Lambda_{03}\simeq1 km/MeV\hspace{3pt}
\end{equation}
\begin{equation}
\label{osc12} Y_{21}=0.8sin^2\Big(\frac{\pi}{\Lambda_{21}}\frac{L
}{E}\Big), \quad  \Lambda_{21}\simeq33 km/MeV\hspace{3pt}
\end{equation}
\begin{equation}
\label{osc13} Y_{31}=0.06sin^2\Big(\frac{\pi}{\Lambda_{31}}\frac{L
}{E}\Big), \quad  \Lambda_{31}\simeq1 km/MeV\hspace{3pt}
\end{equation}
\begin{equation}
\label{osc23} Y_{32}=0.03sin^2\Big(\frac{\pi}{\Lambda_{32}}\frac{L
}{E}\Big), \quad
 \Lambda_{32}\simeq1 km/MeV\hspace{3pt}
\end{equation}
Figure \ref{oscf} displays,  the oscillatory modes $Y_{0i}$ as a
function of $L/E$. As it is seen, the wavelength  of  $Y_{01}$ is
quite large (about 354 km), while the oscillation length  of
$Y_{03}$ is close to that of $Y_{31}$ and $Y_{32}$  ( about 1 km).
The mode $Y_{02}$ has similar oscillation length with that of
$Y_{21}$ mode. Furthermore, the  amplitudes of $Y_{0i}$  modes are
about one to two order of magnitude smaller than   those
  characterizing active neutrino modes
($Y_{21}$,$Y_{31}$, and $Y_{32}$).

\begin{figure}[htb]
\centering
\includegraphics[width=.4\textheight]{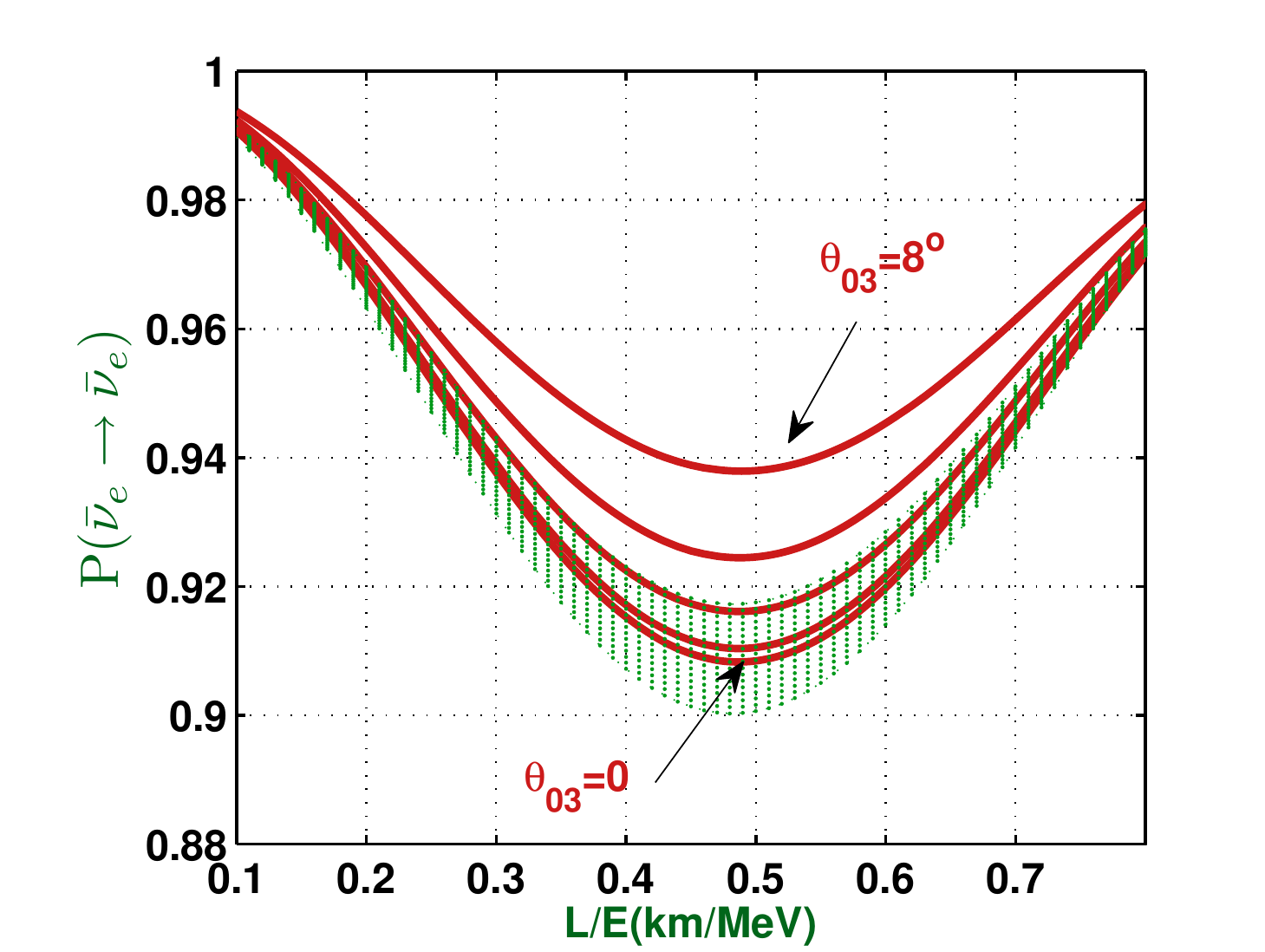}(a)
\includegraphics[width=.4\textheight]{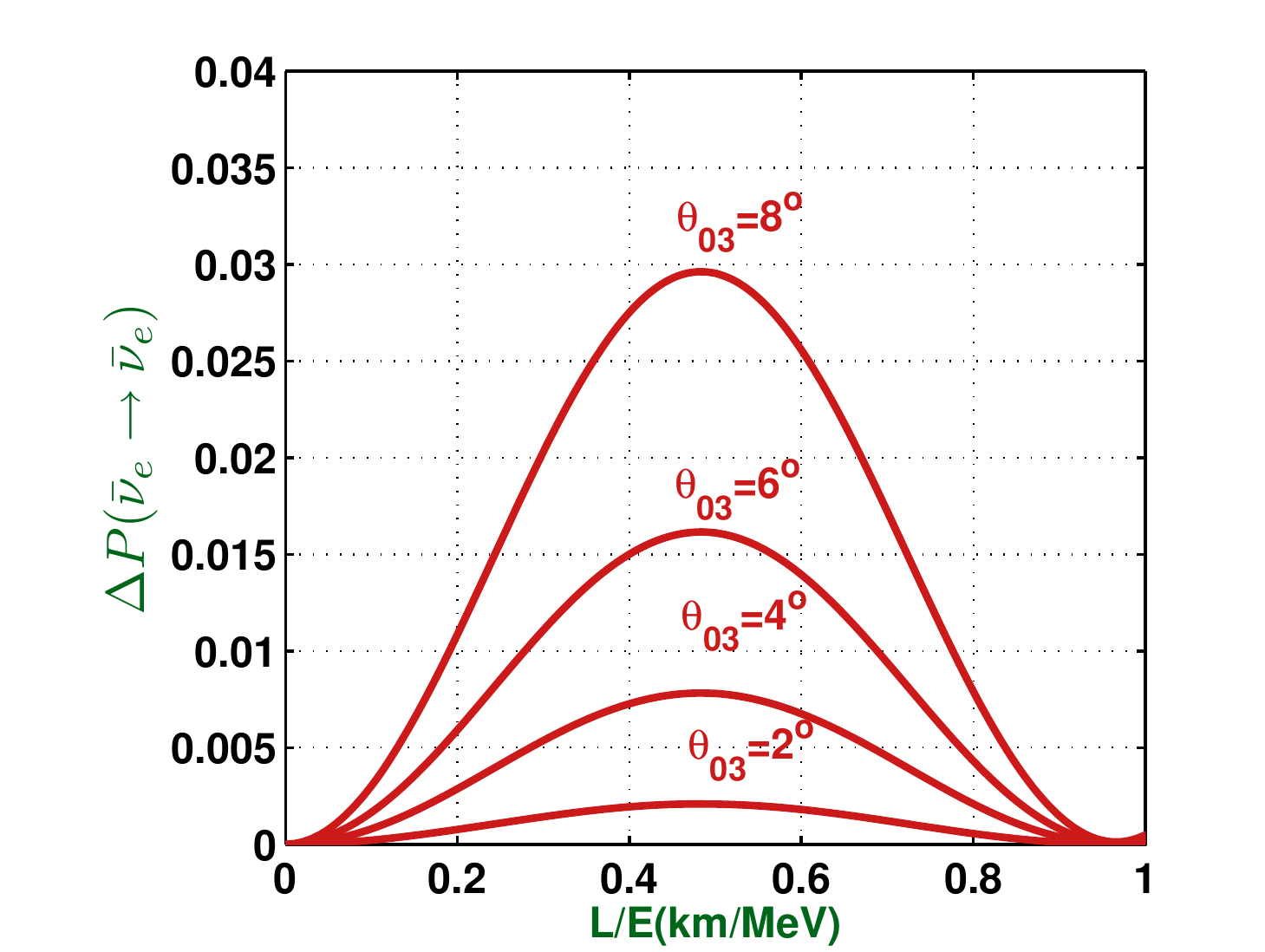}(b)
\caption{(Color on line). (a)  Survival neutrino probabilities
versus $L/E$. Results are given for    sterile mixing angles
$\theta_{03}$ between 0 and  $8^o$ in $2^o$ steps.
The shaded area represents the region of  best fit from the Daya
Bay experiment~\protect\cite{theta13r}. (b) $\Delta P$ difference
of survival neutrino probability in  4$\nu$ model  to   standard
$3\nu$ model as a function of $L/E$ for   mixing angles between
$2^o$ and $8^o$. }
 \label{oscf1}
  \end{figure}

In  Fig.~\ref{oscf1}(a) it is shown  the  survival probability of
4$\nu$-flavor  oscillation model in almost one full oscillation
cycle,  considering various mixing angles  $\theta_{03}$  between
0 and  $8^o$ in $2^o$ steps. As it is seen  results taken with
angle parameter $\theta_{03}$ less than $4^o$ are consistent with
Daya Bay best fit data \cite{theta13r} (shaded area.) Moreover,
the variations induced by the mixing angle $\theta_{03}$ can reach
the level of 3\% around the minimum point (about 0.49 km/MeV) for
$\theta_{03}=8^o$ (see Fig.~\ref{oscf1}(b)). Future experiments
may shed further light on
the allowed intervals  of $\sin^2 2\theta_{0i}$$-$$\Delta
m_{01}^2$,\hspace{2pt} i=1,2,3\hspace{2pt}, oscillation parameters
which could probably be determined with good precision and
sufficient energy resolution, by global fits to future available
experimental $\bar{\nu}_e$$-$disappearance data ( e.g. Ref.
\cite{Bakhti}).

\section{Conclusions}
\label{sec5}

Recent measurements of the energy spectra of the solar neutrino
events at SuperKamiokande, SNO and  Borexino do not show the
expected (according to LMA) upturns at low energies. The absence
of the upturn can be explained by mixing of very light sterile
neutrino in the mass states $\nu_0,\nu_1,\nu_2 $, $\nu_3$
  with mass-squared difference $\Delta m_{01}^2\approx (0.7-2)\times
10^{-5}$ eV$^2$ and mixing angles  $\sin^2 2 \theta_{0i}<
10^{-3}$,\hspace{2pt} $i=1,2,3$. Furthermore, cosmological data,
mainly from observations of the cosmic microwave background and
large scale structure suggest, the existence of a fourth
degree-of-freedom ($N_f>3$) which might be a sterile neutrino.

If   SLSN  exists it  could oscillate with active neutrinos over
the distance of the Earth's radius. A numerical treatment based on
the fourth-order Runge-Kutta method is used to solve the evolution
equation for matter corrections to oscillations of four neutrinos,
adopting a simplified version of the preliminary Earth reference
model.

Taking the neutral-current matter potential $V_n$ to be -0.5 of
the corresponding charge-current $V_e$, we found  a resonant
conversion   at low energies around 17 MeV for   $\Delta m_{01}^2=
0.7\times 10^{-5}$ eV$^2$ and around 37 MeV for $\Delta m_{01}^2=
1.5\times 10^{-5}$ eV$^2$ (neutrino path length $L\approx 12000$
Km). This resonant conversion is much stronger when $V_n=0$ and
becomes much smaller when $V_n$ is taken into account.
Furthermore, when $V_n$ included the resonance position is shifted
to higher energies around 60 MeV, which is  well beyond that of
solar neutrino spectrum. The above results are also sensitive to
mass-squared difference $\Delta m_{01}^2$ as well as to the mixing
angle $\theta_{03}$. It is found that as $\Delta m_{01}^2$
increases the conversion probability amplitude is strongly
suppressed (0.2\% for $\sin^2 2 \theta_{03}=0.005$) with
oscillation pattern to occur in a broadening energy-nadir angle
regions. This makes   difficult to test the scenario of
super-light sterile neutrino  in very low energy atmospheric
neutrino data.

 Furthermore, we have illustrated the effects
induced by the three CP-violating phases through an appropriate
asymmetry, in the energy region of measurable solar neutrinos (0.5
MeV to  20 MeV). We have shown that, the variations induced by the
three unknown CP-violating phases is less than 10\% for the
electron survival probability and can reach the level of 200\% for
the electron-to-sterile transition probability. This significant
departure from zero in the asymmetry could  be interpreted as a
manifestation of leptonic CP violation. If  CP violation occurs
within the context of SLSN model,  then future experiments might
be possible to  point towards a large asymmetry in neutrino
oscillation probability.



It is also interesting to investigate the super light sterile
neutrino scenario  in a medium or  short-baseline
 reactor antineutrino   experiment.
Since  matter effects in a detector are negligible, the
four-neutrino oscillations are based on vacuum-oscillation
solution. It is worth noticing that  the $\bar{\nu_e}$
disappearance exhibits high frequency wiggles in baseline L which
grow smaller as $L$ increases. The ratio of the wiggling period to
the neutrino energy remains  constant. Moreover, the variations of
survival probability  induced by the mixing angle $\theta_{03}$ in
an oscillation length of  $\sim$ 1km  can reach the level of 3\%
around the minimum point (about 0.49 km/MeV) as the mixing
parameter increases.



Nuclear reactors will continue to help us uncover more features
about neutrinos. By enlarging the detector size and/or having more
numerous and powerful sources and/or prolonging the data taking
period, the sensitivity to SLSN can be increased to a desired
level. In the next 20 years, the upcoming next generation reactor
experiments will tell us whether or not super-light sterile
neutrinos exist.




\end{document}